\documentclass[aps,prd,twocolumn,nofootinbib,preprintnumbers,superscriptaddresss,showkeys,showpacs,amsmath,amssymb,include graphics]{revtex4-1}

\usepackage{graphicx}
\usepackage{amsmath,amssymb}
\usepackage{amsfonts}
\usepackage{xspace} 
\usepackage[usenames]{color}
\usepackage{dcolumn}
\usepackage{bm}
\usepackage{mathrsfs}
\usepackage[colorlinks=true]{hyperref}
\usepackage[all]{hypcap} 
\usepackage[utf8]{inputenc} 
\usepackage{slashed}
\usepackage{multirow}
\usepackage{rotating}
\usepackage{nicematrix}
\usepackage{makecell}
\usepackage[bbgreekl]{mathbbol}
\usepackage{relsize}
\DeclareSymbolFontAlphabet{\mathbb}{AMSb}
\DeclareSymbolFontAlphabet{\mathbbl}{bbold}

\begin{document}

\title{Asymptotically conformal CFL quark matter within a nonlocal chiral quark model
}

\author{Oleksii Ivanytskyi$^{1}$} \email{oleksii.ivanytskyi@uwr.edu.pl} 


\affiliation{$^{1}$Incubator of Scientific Excellence---Centre for Simulations of Superdense Fluids, University of Wrocław, Max Born place 9, 50-204, Wrocław, Poland}

\date{\today}

\begin{abstract}

We propose a three-flavor nonlocal NJL model of quark matter with the scalar attractive, vector repulsive and diquark pairing interaction channels.
The model is treated within the separable approximation to obtain the EoS of cold quark matter. 
The analysis of the high density asymptotics of the model allows us to conclude about its qualitative agreement with the perturbative QCD.
Particularly, a color superconducting CFL state is found to be the ground one at asymptotically high densities.
The conformal limit of speed of sound and dimensionless interaction measure are also shown to be reached from below and above, respectively.
The model is applied to modelling NSs within the scenario of early quark deconfinement triggered by the gravitational instability of the NS matter due to the BEC of a spin-color-flavor singlet three-diquark bound state, the light sexaquark, stable against the weak and strong decays. 
A conservative estimate of the sexaquark mass based on the most up-to-date results from QCD sum rules and a constituent quark model implies an early quark deconfinement with the onset mass below $1~\rm M_\odot$.
The scenario is shown to be consistent with the present observational constraints on the mass-radius relation and tidal deformability of NSs.
Given the fact that the proposed model by construction exhibits an asymptotically conformal behavior, we consider the question about approximately conformal quark matter in NSs. 
We report a non-perturbative energy density range, which is inaccessible in NSs and where the speed of sound and dimensionless interaction measure of the CFL quark matter simultaneously attain the conformal values.

\end{abstract}
\maketitle

\section{Introduction}
\label{intro}

Gauge invariance of quantum chromodynamics (QCD) implies an order-by-order agreement of the perturbative calculations within the Coulomb and Lorentz-covariant gauges \cite{Zwanziger:1998ez}.
At the same time the Coulomb gauge has an advantage of dealing exclusively with the physical degrees of freedom. 
This becomes crucial at the non-perturbative regimes where the mentioned agreement is spoiled by the interactions of the longitudinal gluons.
The Coulomb gauge also provides a renormalization-group invariance of the time-time component of the gluon propagator, thus, making it a physically relevant object.
The time-time gluon propagator is also responsible for generating a long-range confining interactions between quarks \cite{Gribov:1977wm,Zwanziger:1995cv}.
Within the Coulomb gauge it simultaneously allows constructing a non-perturbative Hamiltonian, which can be expanded in a quasiparticle basis spaning the Fock space \cite{Szczepaniak:2001rg}.
This gives the access to the quasiparticle representation of QCD and justifies various non-perturbative methods for the low-energy dynamics of quarks including Schwinger-Dyson equations \cite{PhysRevD.33.1785,Roberts:1994dr,Alkofer:2000wg,Watson:2006yq,Binosi:2009qm} and effective Lagrangian approahes rooted in the Nambu–Jona-Lasinio (NJL) model for quark matter \cite{Vogl:1991qt,Klevansky:1992qe,Hatsuda:1994pi,Buballa:2003qv}.

Unlike the effective quark-quark interactions truncated to the processes mediated by the time-time gluon propagator (see e.g. Ref. \cite{Szymanski:2023hvh}), the NJL-type approaches extend beyond the vector interaction kernel in agreement with the analysis of spin splittings in heavy quarkonia \cite{Schnitzer:1975ux}.
Fierz transformation of the (massive) vector boson exchange mode also suggests an argument in favor of existence of various channels of the effective quark-quark interaction \cite{Buballa:2003qv}. 
This makes the NJL model phenomenologically attractive for modelling the dynamics of quark matter at the regimes typical for heavy ion collisions \cite{Vogl:1991qt,Klevansky:1992qe,Hatsuda:1994pi,General:2000zx,Buballa:2003qv,GomezDumm:2001fz,GomezDumm:2004sr,GomezDumm:2005hy,Ratti:2006wg,CamaraPereira:2020rtu,Maslov:2023boq} and interiors of neutron stars (NSs) \cite{Buballa:2003qv,Duhau:2004pq,deCarvalho:2015lpa,CamaraPereira:2016chj,Baym:2017whm,Alvarez-Castillo:2018pve,Song:2019qoh,Ivanytskyi:2019ojt,Contrera:2022tqh,Blaschke:2022egm,Carlomagno:2023nrc}.

However, the NJL model suffers a conceptual problem related to its non-renormalizability manifested through a momentum cut off defining its applicability range.
This drawback is connected to the contact character of the quark-quark interactions.
At the same time, the lattice formulation of QCD (lQCD) indicates that quark interactions act over a certain range in the momentum space \cite{Becirevic:1999kb,Skullerud:2000un,Skullerud:2001aw,Parappilly:2005ei,Blossier:2010vt,Burgio:2012ph,Oliveira:2018lln}.
This implies an additive momentum-dependent mass renormalization in the quark propagator \cite{Skullerud:2001aw}, which is also reported within the Hamiltonian formalism of QCD in Coloumb gauge \cite{Zwanziger:1995cv,Zwanziger:1998ez,Szczepaniak:2001rg,Reinhardt:2017pyr,Quandt:2018bbu}. 
This, in turn, means a space-time nonlocality of the effective quark interactions \cite{Blaschke:1994px,General:2000zx,GomezDumm:2001fz,GomezDumm:2004sr,Duhau:2004pq,GomezDumm:2005hy,Hell:2009by}.
Such a nonlocality also arises within the instanton liquid model \cite{Schafer:1996wv} and the Schwinger-Dyson approach \cite{PhysRevD.33.1785,Roberts:1994dr,Alkofer:2000wg,Watson:2006yq,Binosi:2009qm}.
Thus, nonlocality, is an important feature of the quark-quark interactions that should be accounted for within the effective approaches of the NJL-type by introducing non-contact correlations between quark currents.
Remarkably, this naturally solves the problem of non-renormalizability of the NJL model, since the momentum cutoff mentioned above is not required in this case \cite{Langfeld:1996rn}.

While spontaneous breaking of chiral symmetry of low-energy QCD is incorporated to the nonlocal NJL model by connecting the dynamically generated mass gap to the chiral condensate \cite{Scarpettini:2003fj,Blaschke:2007np,Hell:2008cc,Hell:2009by}, application of the model to the regime of high baryon densities requires accounting for the vector repulsive and diquark pairing interactions \cite{Buballa:2003qv,Baym:2017whm,Song:2019qoh}.
The last two interaction channels were considered within the two-flavor nonlocal NJL model, however, with a drawback of either ignoring the diquark pairing \cite{Plant:1997jr} or treating the vector repulsion locally \cite{Contrera:2022tqh,Carlomagno:2023nrc,Carlomagno:2023nrc}.
The previous versions the three-flavor nonlocal NJL model \cite{deCarvalho:2015lpa,Blaschke:2022egm} have the same drawbacks cured in this work.
We present a nonlocal three-flavor NJL model with vector repulsion and diquark pairing.
We also apply it to modelling the color-superconducting color-flavor-locked (CFL) phase of quark matter \cite{Alford:1997zt,Alford:1998mk,Schafer:1999jg,Alford:2007xm}, which is expected to be the ground state of dense QCD \cite{Schafer:1999jg,Son:1998uk}.
For the sake of simplicity we consider the case when light and strange quarks are degenerated in mass. 
This corresponds to the so-called CFLL quark matter \cite{Blaschke:2023}.
Based on a simple kinematic criterion of {\it "unlocking"} strange quarks from the CFL phase if their mass is high enough \cite{Schafer:1999pb,Alford:1999pa,Abuki:2003ut}, which corresponds to the two-flavor color-superconductivity in quark matter (2SC) \cite{Buballa:2003qv}, we argue that such unlocking does not happen for the considered strengths of the diquark pairing interaction and that the CFLL picture describes dense QCD matter with a reasonable accuracy.
Later we use the developed approach to model NSs with color-superconducting quark cores. 
It is worth mentioning that quark pairing beyond the 2SC and CFL patterns has been shown to be important for the phenomenology of these astrophysical objects \cite{Grigorian:2004jq,Popov:2005xa,Kojo:2020ztt}.
The color-spin locking (CSL) suggests a mechanism of a single-flavor color-superconductivity, which is able to explain such pairing.
The CSL mechanism is applicable even in the case of a substantial Fermi momentum mismatch among light and strange quark flavors caused by the large mass of the latter \cite{Schafer:2000tw,Schmitt:2003xq,Marhauser:2006hy,Aguilera:2006cj,Feng:2009vt}.

Quark deconfinement at high densities is inevitable due to the asymptotic freedom of QCD, which unbinds hadrons \cite{Gross:1973id,Politzer:1973fx}.
But the open questions are whether the deconfinement onsets at the densities reached in the interiors of NSs and what is the corresponding onset mass \cite{Baym:2017whm}?
In the case when the latter is below one solar mass we refer to an early quark deconfinement in NSs.
Recent constraints on their maximum mass well above $2{\rm M}_\odot$ along with the corresponding radius \cite{Miller:2021qha,Riley:2021pdl} exceeding the one of a canonical NS with the mass $1.4{\rm M}_\odot$ \cite{Dietrich:2020efo} are challenging for description within purely hadronic scenarios.
In the presence of hyperons these scenarios also come into a tension with the reported high mass of the black widow pulsar PSR J0952-0607 \cite{Romani:2022jhd}, which potentially provide a more stringent constraint on the NS maximum mass that, nevertheless, should be taken cautiously due to the unknown angular momentum of this accreting system.
On the other hand, this pulsar can be naturally explained as a NS with a color-superconducting quark matter core \cite{Gartlein:2023vif}.
Formation of such cores in the NS interiors generates a characteristic back-bending of the mass-radius relation \cite{Alvarez-Castillo:2018pve,Contrera:2022tqh,Ivanytskyi:2022oxv,Ivanytskyi:2022bjc,Carlomagno:2023nrc,Blaschke:2023,Gartlein:2023vif,Li:2024lmd} providing a good agreement with the mentioned observational constraints.
Such back-bending also leads to relatively small radii of NSs with mass about $1\rm M_\odot$.
This makes the scenario of hybrid quark-hadron NSs consistent with the enigmatic HESS J1731-347 compact object \cite{Sagun:2023rzp}.
The model agnostic statistical analysis of the observational constraints on the NS mass-radius diagram also reports existence of quark matter in their interiors \cite{Annala:2019puf,Annala:2023cwx}.
Finally, quark deconfinement during the supernova explosion suggests a unique mechanism of producing NSs with masses above the two solar ones  \cite{Fischer:2017lag}.
Based on this, the paper considers the scenario of an early quark deconfinement, which has been argued to be prefered by the observational data \cite{Alvarez-Castillo:2018pve,Alvarez-Castillo:2020nkp,Ivanytskyi:2022oxv,Ivanytskyi:2022bjc,Ivanytskyi:2022wln,Blaschke:2022egm,Blaschke:2022knl,Contrera:2022tqh,Sagun:2023rzp,Gartlein:2023vif}.
A phenomenological consequence of this is that all the observed NSs have a quark core.
It is worth mentioning, this possibility is also supported by the recent study of the HESS J1731-347 nature \cite{Sagun:2023rzp}.
As a possible trigger of an early deconfinement in NSs, we discuss the recently proposed Bose-Einstein condensation (BEC) of a color-spin-flavor singlet sexaqauark state causing a mechanical instability of hadron matter against the gravitational compression in NSs and onset of the CFL quark matter \cite{Blaschke:2022knl,Shahrbaf:2022upc}, which.
Nevertheless, thim mechsnism is not build into the model. 

Weakening the quark-quark interactions in the perturbative domain drives the QCD matter toward the conformal limit \cite{Kurkela:2009gj,Fraga:2013qra,Gorda:2018gpy,Fernandez:2021jfr}.
This stimulated intense discussions of the possible nearly conformal behavior of the NS matter as a signature of quark deconfinement 
\cite{Fujimoto:2022ohj,Altiparmak:2022bke,Marczenko:2022jhl,Brandes:2022nxa,Brandes:2022nxa,Takatsy:2023xzf,Annala:2023cwx}.
The nonlocal NJL model presented below is asymptotically conformal by construction.
This enables using it to test how close to the conformal behavior is the QCD matter at the densities reached in the centers of the heaviest NSs.
Below we address this question and demonstrate the existence of a non-perturbative range of densities inaccessible in NSs, where the speed of sound and dimensionless interaction measure of cold quark matter simultaneously attain nearly conformal values.


The paper is organized as follows. 
A nonlocal NJL model for three flavor color-superconducting quark matter with vector repulsion is introduced in the next section.
In Section \ref{sec3} we consider the high density behavior of the model and demonstrate its agreement with the perturbative QCD.
Reaching the conformal limit of cold QCD is also discussed in that section.
Section \ref{sec4} is devoted to modelling NSs with quark cores within the scenario of quark deconfinement driven by the BEC of sexaquarks in cold and dense medium.
That section also discusses the assumption about a nearly conformal behavior of the NS matter.
The conclusions are given in Section \ref{concl}.

\section{Nonlocal NJL model for CFLL quark matter}
\label{sec2}

The NJL model suggests a low energy approximation of QCD and reduces the complexity of quark interactions mediated by gluons to the current-current form.
Within the separable approximation \cite{Blaschke:1994px,GomezDumm:2001fz} the nonlocality of quark interactions is absorbed to the space-time dependent formfactor $g(z)$ defining the quark currents responsible for various interaction channels.

The present consideration is limited to the interaction channels, which acount for the most important aspects of the phenomenology of cold and dense quark matter.
Dynamical restoration and breaking of chiral symmetry and generation of the quark mass gap is among them.
The minimal scheme for accounting for this aspect requires the scalar interaction channel, which includes eight scalar current densities labeled by the index $a=0,8$
\begin{eqnarray}
\label{I}
s_a(x)=\int dz~g(z)\overline q\left(x+\frac{z}{2}\right)\tau_aq\left(x-\frac{z}{2}\right).
\end{eqnarray}
These scalar current densities are defined through the three-flavor quark field $q=(u,d,s)^{T}$.
The integration here is performed over the four-dimensional space-time of the four-volume $\int dz=\beta V$ given in terms of the spacial one $V$ and inverse temperature $\beta=1/T$.
The Gell-Mann matrices $\tau_{a\ge1}$ in the previous equation are related to the generators of the corresponding SU(3) flavor group, while $\tau_0=\sqrt{2/3}$ is introduced for the sake of unifying the notations.
It is worth mentioning, that chirally symmetric formulation of the interaction responsible for the dynamical restoration or breaking of chiral symmetry requires also the pseudoscalar current densities, which can be obtained by replacing $\tau_a$ in Eq. (\ref{I}) by $i\gamma_5\tau_a$.
However, within the mean-field approximation (MFA) considered below introduction of these pseudoscalar current densities does not lead to any observable effects since the corresponding expectation values vanish.
Thus, they are omitted for the sake of simplicity of the consideration. 

The repulsive interaction in dense quark matter motivated by the non-perturbative gluon exchange \cite{Song:2019qoh} is needed for stiffness of quark equation of state (EoS). 
This enables NSs with quark cores to reach two solar masses \cite{Antoniadis:2013pzd,NANOGrav:2019jur,Miller:2021qha,Riley:2021pdl}.
Within the nonlocal NJL model such a repulsion is generated by the vector interaction channel given in terms of the vector current density
\begin{eqnarray}
\label{II}
j_\mu(x)=\int dz~g(z)\overline q\left(x+\frac{z}{2}\right)\gamma_\mu q\left(x-\frac{z}{2}\right).
\end{eqnarray}

Any Fermi system with an arbitrary weak attractive interaction at sufficiently high densities is a subject to a Cooper instability \cite{Cooper:1956zz}.
In dense quark matter this leads to formation of a color-superconducting state consituted by the BEC of diquarks \cite{Alford:1997zt,Alford:1998mk,Schafer:1999jg,Alford:2007xm}.
In order to account for this we introduce to the consideration the diquark current density
\begin{eqnarray}
\label{III}
d_{ab}^{ }(x)=\int dz~g(z)\overline q\left(x+\frac{z}{2}\right)i\gamma_5\tau_a\lambda_bq^c\left(x-\frac{z}{2}\right).
\end{eqnarray}
Here $\lambda_b$ are Gell-Mann matrices related to the generators of the SU(3) color group, the indexes $a,b=2,5,7$ correspond to the antisymmetric color-flavor operators, the charge conjugate quark field is $q^c=i\gamma_2\gamma_0\overline{q}^T$ \footnote{An extended discussion of the diquark operators can be found in the review \cite{Buballa:2003qv}.}.
It is worth noting that the current (\ref{III}) corresponds to condensation of scalar diquarks, which are favored by QCD \cite{Hess:1998sd,Alexandrou:2006cq,Babich:2007ah,DeGrand:2007vu}, while the axial-vector realization of the diquark ground state (see e.g. Ref. \cite{Anselmino:1992vg}) is not considered in this paper.

Eqs. (\ref{I}) - (\ref{III}) do not specify space-time dependence of the formfactor $g(z)$.
The form used in this work is discussed below.
As is seen, at $g(z)=\delta(z)$ the scalar, vector and diquark current densities converge to the operators of the local NJL model \cite{Buballa:2003qv,Baym:2017whm}.
At the same time, the Lagrangian of the NJL model for any particular choice of the formfactor reads
\begin{eqnarray}
\mathcal{L}&=&\overline{q}(i\slashed\partial-m+\mu\gamma_0)q\nonumber\\
\label{IV}
&+&G_S\hspace*{-.1cm}\sum_{a=\overline{0,8}}\hspace*{-.1cm}s_as_a-G_Vj_\mu j^\mu
+3G_D\hspace*{-.3cm}\sum_{a,b=2,5,7}\hspace*{-.3cm}d_{ab}^+d_{ab}^{ }.
\end{eqnarray}
Here $m$ is the current quark mass.
For the CFLL quark matter considered in this work this mass is the same for three quark flavors.
In the case of such a mass degeneration densities of all the quark flavors coincide, while color and electric charge neutrality is provided by a common quark
chemical potential $\mu$ \cite{Blaschke:2022egm}.
The $\beta$-equilibrium is also authomatically provided in this case.
The quark chemical potential enters the Lagrangian (\ref{IV}) via the term including the operator $\overline{q}\gamma_0q$ of conserved quark number density.
The scalar, vector and diquark couplings $G_S$, $G_V$ and $G_D$ control the strength of interaction in the scalar, vector and diquark channels discussed above. 
The conventional factor $3$ in the last term in Eq. (\ref{IV}) is introduced for the sake of further convenience.
Following the commonly accepted practice, in this work we parameterize the vector and diquark couplings via the dimensionless parameters $\eta_V=G_V/G_S$ and $\eta_D=G_D/G_S$.

The first step in treating the present model corresponds to the bosonization of its Lagrangian.
For this we generalize the procedure used in Refs. \cite{Blaschke:2013zaa,Ivanytskyi:2022oxv} and introduce the collective scalar $\sigma_a$, vector $\omega_\mu$ and diquark $\Delta_{ab}^{ }$, $\Delta_{ab}^*$ fields coupled to the current density operators $s_a$, $j_\mu$, $d_{ab}^+$ and $d_{ab}^{ }$, respectively.
They appear in the consideration via the identical Habbard-Stratanovich transformation of the partition function
\begin{eqnarray}
\mathcal{Z}&=&\int\mathcal{D}\overline{q}\mathcal{D}q\exp\left(\int dx~\mathcal{L}\right)\nonumber\\
\label{V}
&=&\int\mathcal{D}\overline{q}\mathcal{D}q\mathcal{D}\sigma_a
\mathcal{D}\omega_\mu\mathcal{D}d_{ab}^*\mathcal{D}d_{ab}^{ }
\exp\left(\int dx~\mathcal{L}^{\rm bos}\right),\quad
\end{eqnarray}
where the bosonized Lagrangian is
\begin{eqnarray}
\mathcal{L}^{\rm bos}\hspace*{-.1cm}&=&\overline{q}(i\slashed\partial-m+\mu\gamma_0)q
\nonumber\\
&-&\sum_{a=\overline{0,8}}\left(s_a\sigma_a+\frac{\sigma_a\sigma_a}{4G_S}\right)
+j_\mu\omega^\mu+\frac{\omega_\mu\omega^\mu}{4G_V}\nonumber\\
\label{VI}
&-&\sum_{a,b=2,5,7}\hspace*{-.2cm}
\left(\frac{d_{ab}^+\Delta_{ab}^{ }+\Delta_{ab}^*d_{ab}^{ }}{2}+
\frac{\Delta_{ab}^*\Delta_{ab}^{ }}{12G_D}\right).
\end{eqnarray}
This bosonized Lagrangian gives a direct access to the Euler-Lagrange equations for the bosonic fields
\begin{eqnarray}
 \label{VII}
 \sigma_a&=&-2G_Ss_a,\\
 \label{VIII}
 \omega_\mu&=&-2G_V j_\mu,\\
 \label{IX}
 \Delta_{ab}&=&-6G_Dd_{ab}.
\end{eqnarray}
The corresponding equation for the $\Delta_{ab}^*$ field can be obtained as the hermitian conjugate of Eq. (\ref{IX}).

From now on we want to sharpen the consideration on the MFA. 
In this case functional integrations over the bosonic fields in Eq. (\ref{V}) are suppressed, while the fields themselves are replaced by their expectation values.
The latter can be found by averaging Eqs. (\ref{VII})-(\ref{IX}).
The scalar currents generated by the non-diagonal flavor Gell-Mann matrices do not survive the averaging procedure due to the vanishing of the corresponding functional integrals over the quark fields.
This conclusion can be independently drawn from the condition of charge conservation provided by the commutation of the mean-field Lagrangian with the charge matrix $\hat Q = {\rm diag}(2/3,-1/3,-1/3)$ \cite{Hell:2009by}.
This requires the absence of the non-diagonal flavor group generators in the mean-field Lagrangian, i.e. vanishing of the corresponding scalar currents.
Furthermore, flavor degeneration in the CFLL quark matter along with the traceless character of $\tau_3$ and $\tau_8$ leads to vanishing of the expectation values of the scalar current densities $s_3$ and $s_8$.
Consequently, only the scalar field $\sigma_0$ attains a non-vanishing value within the MFA to the present model, i.e. $\langle\sigma_a\rangle=\sigma\delta_{a0}$ with
\begin{eqnarray}
 \label{X}
 \sigma=-2G_S\langle s_0\rangle.
\end{eqnarray}
The vector field average can be brought to the form $\langle\omega_\mu\rangle=\omega g_{\mu0}$ by means of the proper Lorentz transformation.
Below the constant $\omega$ is referred to as the vector field. 
It is obtained by averaging Eq. (\ref{VIII}) as
\begin{eqnarray}
 \label{XI}
 \omega=-2G_V\langle j_0\rangle.
\end{eqnarray}
Locking of the color and flavor indexes of quarks in the CFL matter means that only three diquark condensates survive the averaging in the ground state \cite{Pisarski:1998nh}.
It is possible to define a color-flavor rotation under which they are the condensates generated by the antisymmetric SU(3) flavor and color generators with $a=b$ \cite{Buballa:2003qv}. The same holds for the diquark current densities since they have identical color-flavor structure as the diquark condensates, i.e. $\langle d_{ab}\rangle =\langle d_{aa}\rangle\delta_{ab}$.
In the CFLL phase all the quark color-flavor states are equivalent making three non-vanishing diquark condensates equal.
This allows us to introduce
\begin{eqnarray}
 \label{XII}
 \Delta=-2G_D\sum_{a=2,5,7}\langle d_{aa}\rangle.
\end{eqnarray}

Using the above arguments the bosonized Lagrangian (\ref{VI}) can be brought to the mean-filed form. 
This mean-field Lagrangian can be utilized within the Nambu-Gorkov formalism in order to integrate out the quark fields in Eq. (\ref{V}) and arrive at the thermodynamic potential $\Omega=-\ln\mathcal{Z}/\beta V$ \cite{Buballa:2003qv,Blaschke:2013zaa}.
This yields
\begin{eqnarray}
\label{XIII}
\Omega=-\frac{1}{2\beta V}{\rm Tr}\ln\left(\beta\mathcal{S}^{-1}\right)+
\frac{\sigma^2}{4G_S}-\frac{\omega^2}{4G_V}+\frac{|\Delta|^2}{4G_D}.
\end{eqnarray}
Here the trace operation is performed over the Nambu-Gorkov, Dirac, color, flavor, three-momentum and Matsubara indexes, while the inverse Nambu-Gorkov Propagator in the momentum representation is
\begin{eqnarray}
\label{XIV}
\mathcal{S}^{-1}=\left(
\begin{array}{ll}
\hspace*{.6cm}S^{-1}_+\hspace*{.7cm}i\Delta^{ }g_k\gamma_5\mathcal{O}\\
i\Delta^*g_k\gamma_5\mathcal{O}\hspace*{.9cm}S^{-1}_-
\end{array}\right).
\end{eqnarray}
The zeroth component of the four-momentum $k$, which enters the operators $S^{-1}_\pm=\slashed k-M_k\pm\gamma_0(\mu+\omega g_k)$, represents the Matsubara frequencies of fermions, i.e. $k_0=i(2n+1)\pi T$.
The operator $\mathcal{O}=\tau_2\lambda_2+\tau_5\lambda_5+\tau_7\lambda_7$ acts in the color-flavor space and is introduced for shortening the notations.
Following Ref. \cite{Blaschke:2022egm} we introduced the effective momentum-dependent mass $M_k=m+\sigma g_k$.
It is given in terms of the Fourier transformed formfactor $g_k$, which in what follows is referred to as the formfactor.

The vector field enters the Nambu-Gorkov propagator in the combination with the quark chemical potential as $\mu+\omega g_k$.
Following the local formulations of the NJL model with $g_k=1$, this combination is sometimes referred to as the effective momentum-dependent chemical potential of quarks.
We, however, prefer to omit such a notion since it produces a controversy in the context of the grand canonical ensemble.
Indeed, chemical potential is a collective quantity conjugated to a conserved charge and characterizing all the particle excitations carrying that charge.
In the case of a momentum-dependent chemical potential different momentum modes have different effective chemical potentials, which formally splits them to different species with own values of the conserved charge.
This interpretation obviously contradicts the real physical picture when all the momentum modes have the same value of the conserved charge.
Therefore, in this paper we prefer to absorb the vector field term $\omega g_k$ to the single quark energies of quarks.
Mathematically this is equivalent to a momentum-dependent chemical potential mentioned above but does not lead to the discussed controversy in the grand canonical ensemble.

If chosen in a covariant form, $g_k$ explicitly depends on the Matsubara frequencies of fermions entering through the temporal component of the quark four-momentum.
This introduces an additional dependence of the Nambu-Gorkov propagator on the fermion Matsubara frequencies and significantly complicates treating the quark term in Eq. (\ref{XIII}).
In this case the Matsubara summation no longer can be performed explicitly and an additional integration over the temporal component of the quark four-momentum appears \cite{Blaschke:2007ri,Orsaria:2012je,Orsaria:2013hna,Alvarez-Castillo:2018pve}.
This complication is naturally omitted within the instantaneous approximation to quark interaction. 
In this case temporal dependence of the formfactor in the coordinate space is reduced to $\delta(z_0)$, while in the momentum space it depends on the quark three-momentum $\bf k$ only.
In this paper such a non-covariant formfactor is considered.
Below we will use the subscript index "$\bf k$" in order to label the corresponding momentum-dependent quantities. 

Evaluating the quark term in Eq. (\ref{XIII}) involves solving the single particle energies of quarks shifted by the quark chemical potential. 
They are defined as a solution of the equation $\det\mathcal{S}^{-1}=0$ with respect to the temporal component of the quark four-momentum.
In the general case of different current quark masses finding this solution is only possible numerically, while flavor degeneration in the CFLL quark matter allows to do it analytically.
This is equivalent to diagonalizing the Nambu-Gorkov propagator.
The corresponding basis includes singlet and octet states of paired quarks \cite{Shovkovy:1999mr}.
In what follows these states are labeled by the subscript index $j = \rm sing$ and $j = \rm oct$, while their degeneracy factors are $d_{\rm sing}=2\times1$ and $d_{\rm oct}=2\times8$, respectively, where the factors $2$ account for the the spin degrees of freedom.
Thus, the single particle energies of quarks shifted by the chemical potential are
\begin{eqnarray}
\label{XV}
\epsilon^a_{j{\bf k}}={\rm sign}\left(\epsilon_{\bf k}^a\right)
\sqrt{{\epsilon_{\bf k}^a}^2+\Delta_{j\bf k}^2},
\end{eqnarray}
where the superscript index "$a$" labels quarks ($a=+$) and antiquarks ($a=-$), $\epsilon_{\bf k}^\pm=\epsilon_{\bf k}\mp\mu\mp\omega g_{\bf k}$ is expressed through $\epsilon_{\bf k}=\sqrt{{\bf k}^2+M_{\bf k}^2}$ and $\Delta_{j\bf k}=\zeta_j|\Delta|g_{\bf k}$ with $\zeta_{\rm sing}=2$ and $\zeta_{\rm oct}=1$.
It is worth mentioning that both the thermodynamic potential (\ref{XIII}) and the single particle energies (\ref{XV}) include modulus of $\Delta$.
Therefore, for shortening the notations below we suppress the modulus sign when referring to that variable.

It follows from Eq. (\ref{XV}) that the Fermi momentum of quarks $k_F$ is defined by the condition $\epsilon_{\bf k}^+=0$.
Note, for definiteness we consider positive values of the chemical potential.
It is also clear that at this momentum the single particle energy of quarks experiences a discontinuous jump.
Its amplitude $2\Delta_{j\bf k}|_{|{\bf k}|=k_F}$ is twice the gap of the energy spectrum of the corresponding quark state.
Since it is defined by $\Delta$ generated by the diquark pairing, in what follows the latter is referred to as the diquark pairing gap.

Thus, the thermodynamic potential of the CFLL quark matter becomes
\begin{eqnarray}
\Omega&=&-\sum_{j,a}d_j\int\frac{d\bf k}{(2\pi)^3}\left[\frac{\epsilon^a_{j\bf k}}{2}-T\ln\left(1-f^a_{j\bf k}\right)\right]\nonumber\\
\label{XVI}
&+&\frac{\sigma^2}{4G_S}-\frac{\omega^2}{4G_V}+\frac{\Delta^2}{4G_D}.
\end{eqnarray}
Here $f^a_{j\bf k}=\left(e^{\beta\epsilon^a_{j\bf k}}+1\right)^{-1}$ is the single particle distribution function of quarks.
The first term in the quark contribution to $\Omega$ represents a divergent zero point term.
It can be regularized by subtracting a diverging constant being the non-regularized thermodynamic potential of free quarks in the vacuum.
This leads to the regularized thermodynamic potential
\begin{eqnarray}
\label{XVII}
{\Omega^{}}^{\rm reg}&=&\Omega+d\int\frac{d\bf k}{(2\pi)^3}\epsilon^{\rm free}_{\bf k},
\end{eqnarray}
where $d=2\times3\times3$ is the spin-color-flavor degeneracy factor of quarks and $\epsilon^{\rm free}_{\bf k}=\sqrt{{\bf k}^2+m^2}$.

The relation between the partition function and thermodynamic potential along with the expression for the mean field Lagrangian allows us to find the expectation values of the scalar, vector and diquark current densities as
\begin{eqnarray}
\label{XVIII}
\langle s_0\rangle&=&\frac{\partial\Omega}{\partial\sigma}-2G_S\sigma,\\
\label{XIX} 
\langle j_0\rangle&=&\frac{\partial\Omega}{\partial\omega}+2G_V\omega,\\
\label{XX} 
\sum_{a=2,5,7}\langle d_{aa}\rangle&=&\frac{\partial\Omega}{\partial\Delta}-2G_D\Delta.
\end{eqnarray}
The second terms in these expressions compensate the corresponding derivatives of the last three terms in Eq. (\ref{XVI}).
Inserting these expressions to Eqs. (\ref{X}) - (\ref{XII}) we immediately conclude that solutions of the mean-field equations for $\sigma$, $\omega$ and $\Delta$ extremize the thermodynamic potential and can be found from
\begin{eqnarray}
\label{XXI}
\sigma&=&2G_S\sum_{j,a}d_j\int\frac{d\bf k}{(2\pi)^3}\left[\frac{1}{2}-f^a_{j\bf k}\right]\frac{\epsilon^a_{\bf k}}{\epsilon^a_{j\bf k}}
\frac{M_{\bf k}g_{\bf k}}{\epsilon_{\bf k}},\\
\label{XXII}
\omega&=&2G_V\sum_{j,a}d_j^a\int\frac{d\bf k}{(2\pi)^3}\left[\frac{1}{2}-f^a_{j\bf k}\right]\frac{\epsilon^a_{\bf k}}{\epsilon^a_{j\bf k}}
g_{\bf k},\\
\label{XXIII}
\Delta&=&2G_D\sum_{j,a}d_j\int\frac{d\bf k}{(2\pi)^3}\left[\frac{1}{2}-f^a_{j\bf k}\right]\frac{\Delta}{\epsilon^a_{j\bf k}}
\zeta_j^2g_{\bf k}^2.
\end{eqnarray}
Here $d_j^\pm=\pm d_j$ is introduced for shortening the notations.
Unlike the local formulations of chiral quark models (see e.g. Ref. \cite{Ivanytskyi:2022oxv,Ivanytskyi:2022bjc}), the previous equations do not include terms with the Dirac delta-function $\delta(\epsilon^a_{\bf k})$ arising from the derivative of the sign-function in the dispersion relation (\ref{XV}).
In the case of quarks it is provided by $1/2-f^+_{j\bf k}=0$ at the quark Fermi momentum, while for antiquarks the delta-function vanishes itself due to $\epsilon^-_{\bf k}>0$. 
It is worth mentioning that due to the presence of the formafactor under the momentum integrals Eqs. (\ref{XXI}) - (\ref{XXIII}) do not require any additional regularization of the zero point terms. 
Their convergence is automatically provided if at large momenta $g_{\bf k}$ decreases with growth of $\bf k$ faster than ${\bf k}^{-2}$.
It is important to note, this feature also provides the renormalizability of the model since no additional ultraviolet momentum cutoff is needed in this case.

The solutions of Eqs. (\ref{XXI}) - (\ref{XXIII}) can be used in order to obtain the regularized thermodynamic potential, which gives a direct access to the EoS of the CFLL quark matter. 
The pressure defined as $p=\Omega^{\rm reg}_0-{\Omega^{}}^{\rm reg}$ vanishes in the vacuum ($T=\mu=0$).
Below all the vacuum quantities are labeled with the subscript index ''$0$".
The baryon number, entropy and energy densities can be found using the thermodynamic identities $n_B=\partial p/\partial\mu_B$, $s=\partial p/\partial T$ and $\varepsilon=\mu_B n_B+Ts-p$.
Here $\mu_B=3\mu$ is the baryochemical potential.
Below we also analyze the speed of sound defined as the derivative $c_S^2=dp/d\varepsilon$ evaluated at a constant value of the entropy per baryon $s/n_B$.
In the zero temperature regime $s=0$ and this  last condition is fulfilled automatically leading to $c_S^2=\partial\ln\mu_B/\partial\ln n_B$.
We will also consider the dimensionless interaction measure $\delta=1/3-p/\varepsilon$, which is nothing but trace of the energy momentum tensor normalized by $3\varepsilon$.
This quantity is widely used for discussion of the conformal behavior of quark matter. 

This paper is exclusively focused on the cold quark matter, which is relevant for modelling NSs.
Therefore, in what follows $T=0$. 
This reduces the single particle distribution functions of quarks to unit step-functions and provide
\begin{eqnarray}
\label{XXIV}    
T\ln\left(1-f_{j\bf k}^a\right)=\epsilon_{j\bf k}^af_{j\bf k}^a=
\epsilon_{j\bf k}^a\theta\left(-\epsilon_{j\bf k}^a\right).
\end{eqnarray}
In what follows we use the Gaussian formfactor
\begin{eqnarray}
\label{XXV}
g_{\bf k}=\exp\left(-{\bf k}^2/\Lambda^2\right),
\end{eqnarray}
where constant $\Lambda$ defines the momentum scale below which the effective current-current interactions are important. 
This choice guarantees ultraviolet convergence of the present model and its qualitative agreement with perturbative QCD results in reaching the conformal limit of quark matter.
At high densities most of the single quark states have momenta well above $\Lambda$.
This diminishes the interaction effects and drives the system toward the conformal limit discussed in Section \ref{sec3}.

The formfactor can be chosen in other functional forms, e.g. in the Lorentzian-type or Woods-Saxon ones \cite{Grigorian:2006qe}.
However, at non-perturnative regimes typical for the NS interiors specific choice of $g_{\bf k}$ is not too important since $k_F<\Lambda$ and for the momentum states below the Fermi sphere all the mentioned functional forms of the formfactor differ by not more than 25\%.
This makes modelling NS within the framework of the nonlocal NJL model weakly sensitive to the formfactor choice.

Given the simplification related to the mass degeneration of the quark flavors, we aim at finding reasonable values of the model parameters rather than at their ultimate determination.
Thus, the current quark mass is fixed as the mean value of the current masses of $u$- and $d$-quarks from the Review of Particle Physics \cite{ParticleDataGroup:2022pth}, i.e. $m=(m_u+m_d)/2=3.5$ MeV.
Furthermore, $\Lambda$ and $G_S$ are fixed according to the vacuum values of the effective quark mass at vanishing momentum $M_{0,{\bf k}=0}$ and chiral condensate per flavor $\langle\overline{f}f\rangle_0$. 
The latter can be found from the regularized thermodynamic potential as 
\begin{eqnarray}
 \label{XXVI}
 \langle\overline{f}f\rangle_0=\frac{1}{3}\frac{\partial\Omega^{\rm reg}_0}{\partial m}=\frac{d}{3}\int\frac{d\bf k}{(2\pi)^3}
 \left[\frac{m}{\epsilon^{\rm free}_{\bf k}}-\frac{M_{0,\bf k}}{\epsilon_{0,\bf k}}\right].
\end{eqnarray}
Depending on the current quark mass and gauge adopted in the lQCD simulations, the vacuum effective mass at zero momentum is found to lay in the interval 300-500 MeV \cite{Skullerud:2000un,Skullerud:2001aw,Parappilly:2005ei,Burgio:2012ph}.
The QCD sum rules at the renormalization scale of 1 GeV yield $-(229\pm33~\rm MeV)^3$ \cite{Dosch:1997wb} and $-(242\pm15~\rm MeV)^3$ \cite{Jamin:2002ev} for the vacuum chiral condensate per flavor.
The Gell-Mann-Oakes-Renner relation \cite{Gell-Mann:1968hlm} with the pion mass and decay constant from the Review of Particle Physics \cite{ParticleDataGroup:2022pth} and the current quark mass mentioned above predicts $-(280\pm12~\rm MeV)^3$.
In this work we use the compromise values $M_{0,{\bf k}=0}=400$ MeV and $\langle\overline{f}f\rangle_0=-(250~\rm MeV)^3$. 
Fitting them yields $\Lambda=564~\rm MeV$ and $G_S=9.62~{\rm GeV}^{-2}$.

\begin{figure*}[t!]
\label{fig0}
\includegraphics[width=1\columnwidth]{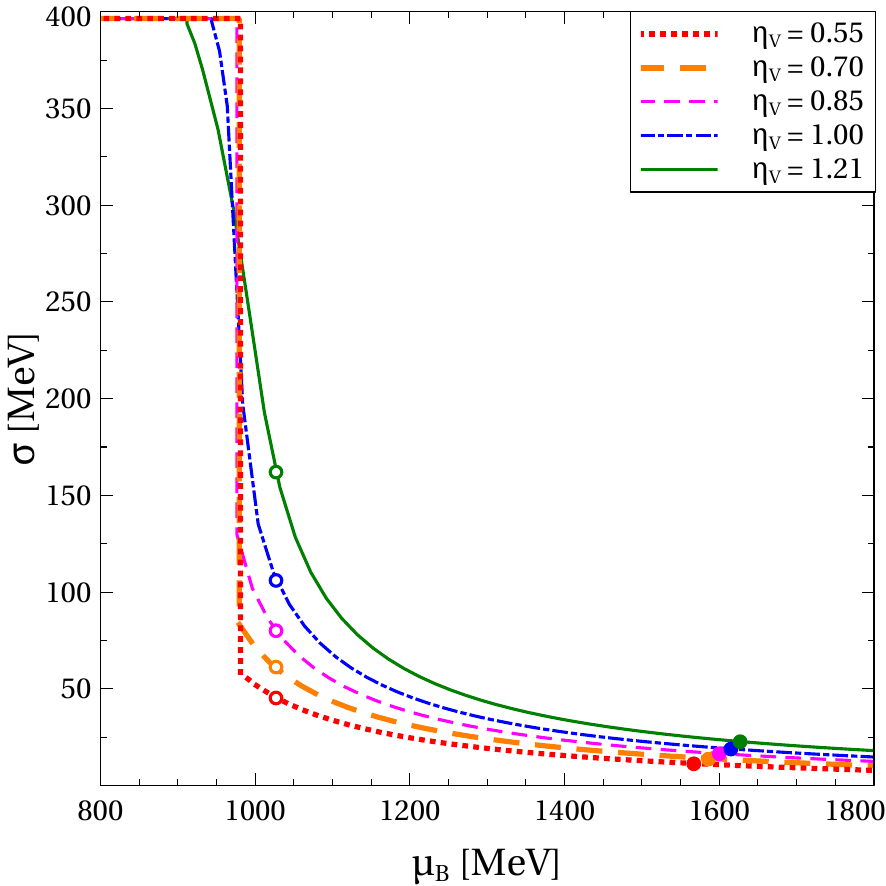}
\includegraphics[width=1\columnwidth]{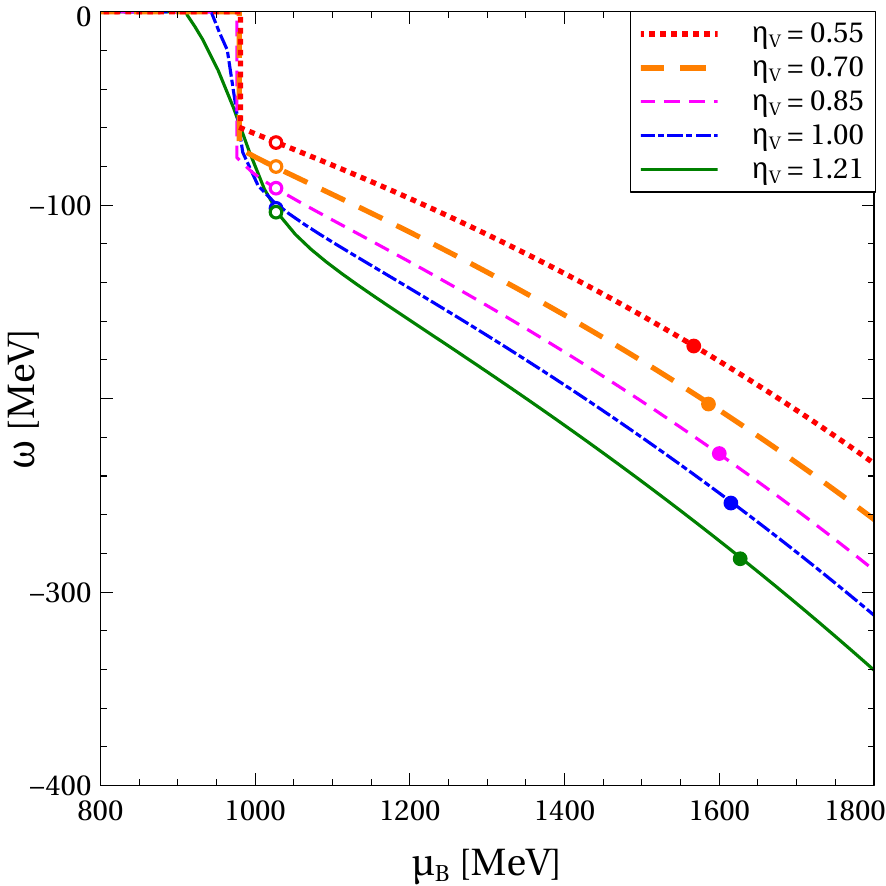}
\includegraphics[width=1\columnwidth]{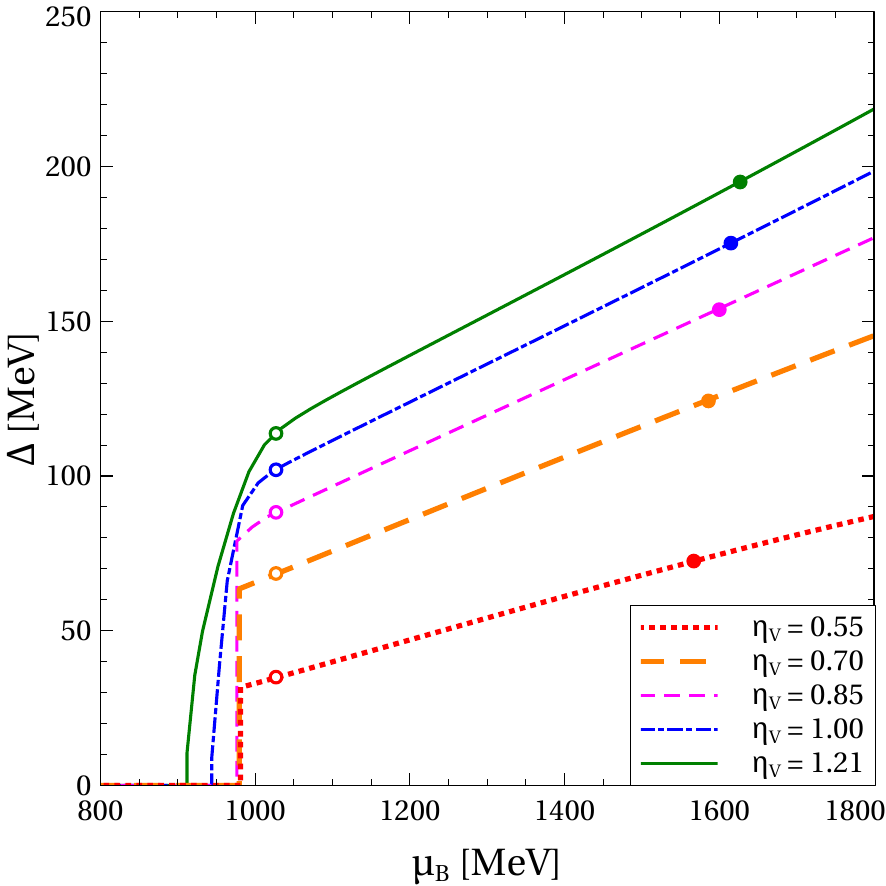}
\includegraphics[width=1\columnwidth]{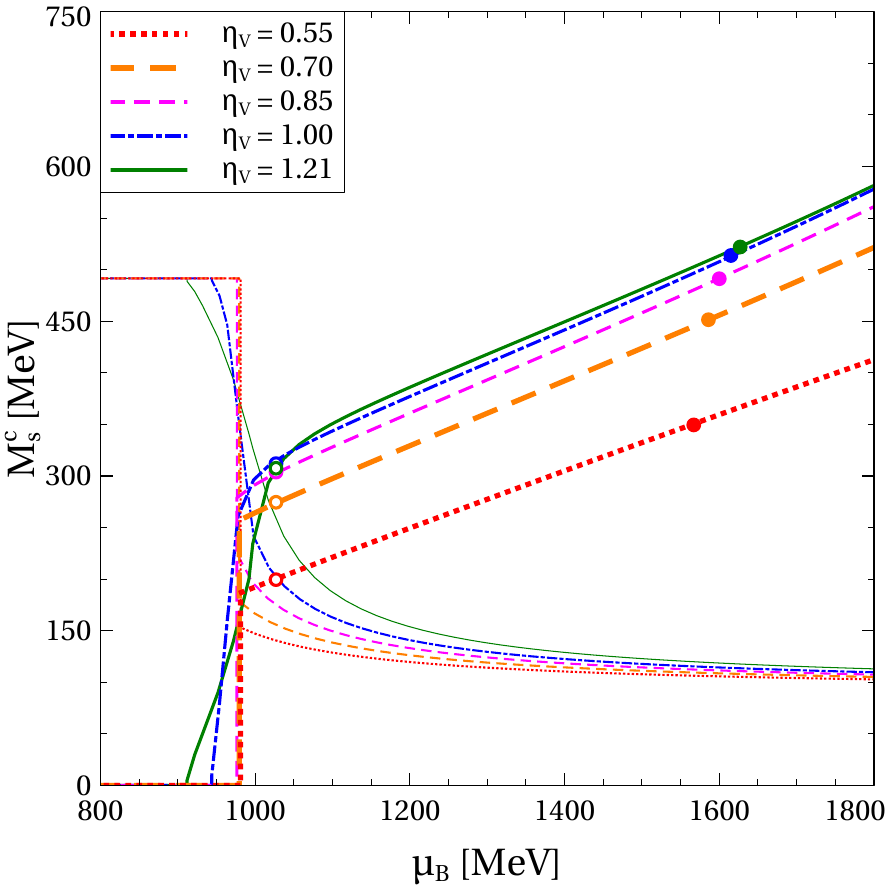}
\caption{
Scalar field $\sigma$ (upper left panel), vector field $\omega$ (upper right panel), pairing gap $\Delta$ (lower left panel) and critical value of the strange quark mass $M_s^c$ as functions of baryon chemical potential $\mu_B$ obtained for the vector and diquark couplings presented in Table \ref{table1}. 
The thin curves on the lower right panel correspond to the effective mass of the zero momentum strange quarks $M_{s,{\bf k}=0}$ calculated at $m_s=95$ MeV.
The empty white circles indicate the first order phase transition in the hybrid quark-hadron EoS (see Section \ref{sec4} for details).
The filled colored circles represent the maximum chemical potentials in the centers of the heaviest NSs with the corresponding EoSs.}
\end{figure*}

We also require the vacuum state to be normal, i.e. not color-superconducting.
The corresponding condition  was found in Ref. \cite{Blaschke:2022egm}.
It reads
\begin{eqnarray}
\label{XXVII}
G_D<\left[\frac{8d}{3}\int\frac{d\bf k}{(2\pi)^3}\frac{g_{\bf k}^2}{\epsilon_{0,\bf k}}\right]^{-1},  
\end{eqnarray}
where the scalar field in the vacuum $\sigma_0=M_{0,{\bf k}=0}-m$ satisfies Eq. (\ref{XXI}) at $T=\mu=\omega=\Delta=0$.
This yields $\eta_D<0.765$.
In what follows $\eta_D$ below the mentioned value and $\eta_V$ are considered as the parameters controlling the high density behavior of the model.

\begin{table}[t]
\centering
\begin{tabular}{|c|c|c|c|c|c|c|}
\hline    
$\eta_V$ & $\eta_D$ & $\rm \varepsilon_{\rm max}$ & $\rm M_{\rm max}$ & $\rm R_{1.4}$ & $\Lambda_{1.4}$ \\
          &         & $\rm [MeV/fm^3]$        & $\rm M_{\rm max}~[M_\odot]$ & $\rm R_{1.4}$ [km] & $\Lambda_{1.4}$ \\ \hline
 0.55  & 0.27 & 1281 & 1.94 & 12.23 & 280 \\ \hline
 0.70  & 0.34 & 1203 & 2.01 & 12.39 & 321 \\ \hline
 0.85  & 0.38 & 1164 & 2.09 & 12.58 & 393 \\ \hline
 1.00  & 0.41 & 1086 & 2.15 & 12.80 & 424 \\ \hline
 1.21  & 0.44 & 1038 & 2.24 & 13.15 & 569 \\ \hline
\end{tabular}
\caption{Vector $\eta_V$ and diquark $\eta_D$ couplings producing the solutions of the mean-field equations shown of Fig. \ref{fig0}, EoSs shown in Fig. \ref{fig1}, maximal energy densities $\varepsilon_{\rm max}$ reached in the centers of the heaviest NSs, as well as the corresponding maximum masses $\rm M_{\rm max}$, radii $\rm R_{1.4}$ and tidal deformabilities $\Lambda_{1.4}$ of a canonical NS extracted from Figs. \ref{fig3} and \ref{fig4}.} 
\label{table1}
\end{table}

Fig. \ref{fig0} shows solutions of the coupled mean-field equations (\ref{XXI}) - (\ref{XXIII}) solved for the values of the model parameters specified above and for the vector and diquark couplings given in Table \ref{table1}.
At small chemical potentials the vector field and pairing gap vanish, while the scalar field has its vacuum value.
This corresponds to the chirally broken normal phase of cold QCD matter, which is the ground state at small $\mu_B$. 
Within the MFA baryon density and pressure of this phase vanish since quarks have large effective masses and can not be excited.
Going beyond the MFA introduces to the consideration nucleons and hyperons, which generate finite baryon densities and pressures even in the absence of quarks.
In Section \ref{sec4} this is modeled by constructing a hybrid quark-hadron EoS.
At high chemical potentials scalar field deviates from its vacuum value and the pairing gap is finite.
This signals about the partially chirally restored color-superconducting phase of cold QCD matter, the CFLL phase.
This phase is the ground state of QCD at high densities and is constituted by the Cooper pair type diquark correlations of quarks with small effective masses.
The vector field, baryon density and pressure attain finite values in this phase.
The transition between the chirally broken normal and partially chirally restored color-superconducting phases of cold QCD matter occurs when their pressures coincide.
As is seen from Fig. \ref{fig0}, depending on the values of the vector and diquark couplings the order parameters of chiral symmetry ($\sigma$) and color superconductivity ($\Delta$) change continuously or discontinuously at this  transition.
At large $\eta_V$ and $\eta_D$ this change is continuous but exhibits kinks in the behavior of $\sigma$ and $\Delta$, which signals about a second order phase transition.
At small values of the vector and diquark couplings the mentioned order parameters change discontinuously and the transition is of the first order.

While the present model neglects the mass splitting between different quark flavors, the quark pairing patterns can be modified if the current mass of strange quarks has a physical value $m_s\gg m$.
More specifically, at low densities heavy strange quarks are not gapped if their effective mass $M_{s,{\bf k}}=m_s+\sigma g_{\bf k}$ exceeds a certain critical value $M_s^c$ \cite{Schafer:1999pb,Alford:1999pa,Abuki:2003ut}.
Here the current quark mass has its physical value $m_s=95$ MeV \cite{ParticleDataGroup:2022pth} and $\sigma$ is calculated at $m_s=m$.
Under the mentioned condition strange quarks are unlocked from the CFL phase of quark matter, which corresponds to the 2SC quark matter \cite{Buballa:2003qv}.
The critical value of the effective mass of strange quarks depends on the strength of the diquark pairing interaction and density of quark matter.
Based on a simple kinematic consideration it reads \cite{Schafer:1999pb,Alford:1999pa,Abuki:2003ut}
\begin{eqnarray}
\label{msc}
M_s^c=2\sqrt{k_F\Delta},
\end{eqnarray}
where the quark Fermi momentum and pairing gap are calculated at $m_s=m$.
Fig. \ref{fig0} shows the behavior of $M_s^c$ and the effective mass of the zero momentum strange quarks $M_{s,{\bf k}=0}$.
Note, at finite momenta $M_{s,{\bf k}}<M_{s,{\bf k}=0}$.
It is seen from Fig. \ref{fig0} that for the considered values of the vector and diquark couplings and at small values of the baryonic chemical potentials below about 1000 GeV $M_{s,{\bf k=}0}>M_s^c$. 
In this range of $\mu_B$ the simplification related to mass degeneration of quarks is inapplicable, since it overlooks the transition between the CFL and 2SC phases of quark matter \cite{Blaschke:2005uj,Blaschke:2008br,Blaschke:2010ka}.
At the same time, $M_s^c$ always exceeds $M_{s,{\bf k}=0}$ above the first order phase transition in the hybrid quark-hadron EoS described in Section \ref{sec4}.
This means that accounting for the physical current mass of strange quarks within the present model does not change the CFL pattern of quark pairing and does not bring it to the 2SC one if $\mu_B$ exceeds about 1000 MeV.
Furthermore, for the vector and diquark couplings providing a good agreement with the observational data on NS ($\eta_V\ge0.7$, see Section \ref{sec4} for details) and $\mu_B>1200$ MeV $M_s^c$ exceeds $M_{s,{\bf k}=0}$ at least by a factor about two.
This allows us to conclude that in the mentioned range of the baryonic chemical potentials accounting for a physical value of $m_s$ produces a small correction to the EoS of quark matter that can be accounted for perturbatively in the spirit of the phenomenological parametrization by Alford, Braby, Paris and Reddy \cite{Alford:2004pf}.

\section{Nonlocal NJL model at high densities}
\label{sec3}

At high densities color superconductivity is the ground state of QCD, while the corresponding pairing gap vanishes only asymptotically \cite{Schafer:1999jg}.
At this regime QCD restores the conformal behavior due to the asymptotic freedom and negligebly small quark masses.
Before going further, we consider the high density behavior of the CFLL quark matter in order to demonstrate that it respects the mentioned features of QCD.
For the sake of generality the following analysis is performed for an arbitrary formfactor.
The Gaussian case is considered after the general expressions for the relevant thermodynamic quantities are obtained.  

First, one has to prove that once a color-superconducting state is formed in the CFLL quark matter, it preserves itself at arbitrary high densities, which is equivalent to the absence of a high density second order phase transition within the present model.
The condition of such a phase transition is $\partial^2\Omega^{\rm reg}/\partial\Delta^2=0$ at vanishing pairing gap.
Dividing Eq. (\ref{XXIII}) by the gap and later setting $\Delta=0$ yields this condition in the explicit form.
The quark contribution to its right hand side part diverges due to the factor $(\epsilon_{\bf k}-\mu)^{-1}$ under the momentum integral.
This divergence is naturally regularised in the presence of a non-zero pairing gap.
Consequently, this right hand side part diverges at vanishing pairing gap and, thus, is not equal to the unity in the left hand side part.
This allows us to conclude that the condition of a second order phase transition at high densities can not be fulfilled and the pairing gap is non-zero.
This proves that, just like in QCD, the color superconductivity is the ground state of cold and dense quark matter.
At the same time, at high densities Eq. (\ref{XXIII}) can not be satisfied by finite or even diverging pairing gap. 
In this case the right hand side part of Eq. (\ref{XXIII}) vanishes at high chemical potentials due to the factor $\epsilon^a_{j\bf k}$ under the momentum integral and, thus, is not equal to $\Delta$ in its left hand side part.
In other words, fulfilling Eq. (\ref{XXIII}) requires $\Delta\rightarrow0$ at $\mu\rightarrow\infty$. 
Thus, the pairing gap of the CFLL quark matter asymptotically vanishes in the high density regime.

This allows us to approximate $\epsilon_{\bf k}^a/\epsilon_{j\bf k}^a\simeq1$ in Eqs. (\ref{XXI}) - (\ref{XXII}), while the subdominant order $\mathcal{O}(\Delta^2)$ correction can be neglected. 
Furthermore, the single particle distribution function of quarks can be decomposed as $f_{j\bf k}^+=1-\theta(|{\bf k}|-k_F)$, where the first and the second terms represent the dominant contribution and the leading order corrections to the momentum integrals. 
Using this decomposition and performing the summation over the quark-antiquark states we obtain the leading terms of the scalar and vector fields
\begin{eqnarray}
 \label{XXVIII}   
 \sigma_\infty&=&2G_Smd\int\frac{d\bf k}{(2\pi)^3}\frac{g_{\bf k}}{|\bf k|}\theta(|{\bf k}|-\mu),\\
 \label{XXIX}   
 \omega_\infty&=&-2G_Vd\int\frac{d\bf k}{(2\pi)^3}g_{\bf k},
\end{eqnarray}
where an obvious relation $d_{\rm sing}+d_{\rm oct}=d$ was used.
Note, in the expression for $\sigma_\infty$ the quark Fermi momentum was replaced by the corresponding chemical potential since the quark effective mass and vector field can be safely neglected compared to $\mu$.
Eq. (\ref{XXVIII}) demonstrates that within the nonlocal NJL model the momentum dependent effective quark mass at high densities converges to the current one, since the scalar field vanishes at this regime.  
At the same time, it follows from Eq. (\ref{XXIX}) that the vector field saturates to a finite value. 

The analysis of the pairing gap should be performed more accurately since at small $\Delta$ the function  $(1/2-f_{j\bf k}^+)/\epsilon_{j\bf k}^+=1/2|\epsilon_{j\bf k}^+|$ has a sharp peak around the quark Fermi momentum and most of the momentum integral comes from the interval of the width $2\delta_\mu$ around the peak.
A $\delta_\mu$ respecting the condition $\Delta\ll\delta_\mu<\mu$ can be chosen arbitrary since, as is shown below, the corresponding contribution to the momentum integral in Eq. (\ref{XXIII}) is subdominant at large chemical potentials. 
Within the mentioned momentum interval the formfactor can be approximated by its value at the momentum equal to the chemical potential.
In what follows it is denoted as $g_\mu$.
The quark mass and vector field can be neglected compared to the chemical potential.
Using this along with $d{\bf k}g_{\bf k}^2\simeq d{\bf k}g_\mu^2\mu^2/{\bf k}^2$, integrating over the mentioned momentum interval and taking the limit of small pairing gap, Eq. (\ref{XXIII}) yields
\begin{eqnarray}
\label{XXX}
1\simeq\frac{G_D\mu^2g_\mu^2}{\pi^2}
\sum_jd_j\zeta_j^2\ln\frac{2\mu\sqrt{\delta_\mu/\mu}}{\zeta_j g_\mu \Delta_\infty}.
\end{eqnarray}
Note, the contribution of antiquarks was neglected due to $\epsilon_{j\bf k}^-\rightarrow\infty$.
At $\delta_\mu\gg\Delta_\infty$ the logarithm in the previous relation can be split to the leading  $\ln(\mu/\Delta_\infty)$ and subdominant $\ln(2\sqrt{\delta_\mu/\mu}/\zeta_j)$ parts. 
Suppressing the subdominant one the pairing gap is expressed as
\begin{eqnarray}
\label{XXXI}
 \Delta_\infty=\frac{\mu}{g_\mu}\exp\left[-\frac{\pi^2}{\tilde dG_D\mu^2g_\mu^2}\right],
\end{eqnarray}
where $\tilde d=d_{\rm sing}\zeta_{\rm sing}^2+d_{\rm oct}\zeta_{\rm oct}^2$. It is interesting to compare this results to the one gluon exchange perturbative calculations \cite{Schafer:1999jg}.
They suggest the quark pairing gap $\Delta_{\rm QCD}\propto\mu g_{\rm QCD}^{-5}\exp(-3\pi^2/\sqrt{2}g_{\rm QCD})$ with $g_{\rm QCD}$ being the QCD coupling. 
Equating $\Delta_\infty$ and $\Delta_{\rm QCD}$ brings the high density behavior of the formfactor $g_\mu\propto \sqrt{g_{\rm QCD}}/\mu$.
Running of the QCD coupling can be estimated according to the one-loop $\beta$-function as $g_{\rm QCD}\propto\ln^{-1/2}(\mu^2/\Lambda^2_{\rm QCD})$.
From this we conclude that agreement with the perturbative QCD calculations requires the formfactor to behave at high momenta as $|{\bf k}|^{-1}$.
This behavior of the formfactor leads to divergent momentum integrals in Eqs. (\ref{XXI}) - (\ref{XXIII}).
At $g_{\bf k}\propto|{\bf k}|^{-1}$ their regularization requires an ultraviolet momentum cutoff, which spoils the renormalizability of the nonlocal NJL model.
Requiring the renormalizability disables using the formafactor extracted from the perturbative QCD calculations of the quark pairing gap.

In the Gaussian case considered in this paper the momentum integrals in Eqs. (\ref{XXVIII}) and (\ref{XXIX}) can be performed explicitly. 
This yields the asymptotes of the vector and scalar fields $\sigma_\infty=dG_S\Lambda^2m\exp(-\mu^2/\Lambda^2)/2$ and $\omega_\infty=-dG_V\Lambda^3/4\pi^{3/2}$.
For the considered formfactor the pairing gap decreases double exponentially at growing $\mu^2$.

Finding the asymptote of the speed of sound and dimensionless interaction measure of cold CFLL matter requires accounting for the fact that quark Fermi momentum is reduced by the vector repulsion and quark mass.
Saturation of the vector field to the constant value found above and convergence of the effective quark mass to the current mass brings us to the conclusion that at high densities $k_F\simeq\mu+\omega_\infty g_\mu-m^2/2\mu$.
For the formfactors providing convergence of the momentum integrals in Eqs. (\ref{XXI})-(\ref{XXIII}) the quark mass correction to $k_F$ dominates against the vector repulsion.
This, however, is not the case in the chiral limit when $m=0$.
Therefore, both the mass and vector repulsion corrections are kept for the sake of generality of the analysis.
At high densities $n_B\simeq dk_F^3/18\pi^2$.
Using this, we arrive at
\begin{eqnarray}
\label{XXXII}
c_S^2\simeq\frac{1}{3}\left[1-\omega_\infty\mu\frac{\partial}{\partial \mu}\frac{g_\mu}{\mu}-\frac{m^2}{\mu^2}\right].
\end{eqnarray}
The asymptote of the dimensionless interaction measure can be found by noticing that at high densities the pressure to energy density ratio is of an indeterminate form $\infty/\infty$, which can be treated using the L’H\^opital's rule. This along with the thermodynamic identities mentioned above yields $\delta=1/3-c_S^2$.
Note that this model independent relation holds only at asymptotically high densities but applies even for the systems, which do not converge to the conformal limit.

Thus, the CFLL quark matter asymptotically approaches the conformal limit $c_S^2=1/3$, $\delta=0$ due to vanishing of the formfactor and negligibly small quark mass.
Such an asymptotically conformal behavior was reported within the confining density functional approach for color-superconducting quark matter with medium dependent couplings \cite{Ivanytskyi:2022bjc,Gartlein:2023vif}.
Within that approach the conformal values of the speed of sound and dimensionless interaction measure are reached from above and below, respectively.
At the same time, the perturbative QCD calculations for cold and dense quark matter indicate reaching them from the opposite sides.
It follows from Eq. (\ref{XXXII}) that given a negative value of $\omega_\infty$, the conformal limit within the present model is reached in a qualitative agreement with perturbative QCD for any formfactor providing convergence of the momentum integrals in Eqs. (\ref{XXI}) - (\ref{XXIII}) regardless the value of $m$.
However, the formfactor can not be chosen to simultaneously provide the renormalizability of the present model and reproduce the two-loop perturbative result assuming that deviation of speed of sound from the conformal value scales as $g_{\rm QCD}^4$ \cite{Blaschke:2023}.
This signals about an inherent non-perturbative character of the nonlocal NJL model.

It is important to stress that the mentioned qualitative agreement of the considered model with the predictions of perturbative QCD is provided by the nonlocal character of the quark-quark interactions.
In order to demonstrate this we want to compare the present mode to the version of the NJL model from Ref. \cite{Blaschke:2022egm}. 
Similarly to this work, that model treats the scalar and diquark interaction channels nonlocally and adopts for them the Gaussian formfactor.
At the same time, the vector interaction of that version of the NJL model is local.
The corresponding thermodynamic potential and mean-field equations can be obtained by replacing $\omega g_{\bf k}$ by $\omega$ where applicable and setting the formafactor in Eq. (\ref{XXII}) equal to unit.
This, however, does not change the conclusion about vanishing $\sigma_\infty$ and $\Delta_\infty$, which holds due to the nonlocal treatment of the corresponding interaction channels.
Then, only the contribution of quarks survives in Eq. (\ref{XXII}) with $g_{\bf k}=1$.
This yields
\begin{eqnarray}
\label{XXXIII}
\omega_\infty=-\frac{G_Vd(\mu+\omega_\infty)^3}{3\pi^2}
\simeq-\mu+\left(\frac{3\pi^2\mu}{G_Vd}\right)^{1/3}.
\end{eqnarray}
Thus, the vector field asymptotically diverges in the local case.
This is in a striking disagreement with the nonlocal NJL model providing a convergent $\omega_\infty$.  
Furthermore, in the local case the quark contribution to the thermodynamic potential scales as $(\mu+\omega_\infty)^4\propto\mu^{4/3}$ and is negligible compared to the contribution of the vector field $-\omega^2/4G_V\propto\mu^2$.
This leads to the pressure $p=\mu^2/4G_V$, speed of sound $c_S^2=1$ and dimensionless interaction measure $\delta=-2/3$ violating the conformal behavior. 

It is also interesting to consider another version of the NJL model with the scalar and vector interactions treated nonlocally but the diquark pairing kept local. 
In this case the formfactor in Eq. (\ref{XXIII}) should be set $g_{\bf k}=1$ and the effects of vector and scalar interactions can be neglected since $\omega_\infty$ is finite and $\sigma_\infty$ vanishes.
As it is shown below, the pairing gap diverges if the corresponding interaction is treated locally.
Moreover, $\Delta_\infty\gg\mu$ in this case, which allows us to approximate $\epsilon^\pm_{j{\bf k}}\simeq\mp\Delta\zeta_j$.
This cancels the zero point terms of quarks and antiquarks in Eq. (\ref{XXIII}) and, unlike the nonlocal NJL model, yields the divergent $\Delta_\infty=G_D\tilde{d}\mu^3/3\pi^2$.
As expected, quark chemical potential is negligibly small compared to this asymptotote of the quark pairing gap.
In this case the quark contribution to the thermodynamic potential reads $-\tilde d\Delta_\infty\mu^3/6\pi^2$, while the resulting pressure, speed of sound and dimensionless interaction measure attain the nonconformal values $p=G_D\tilde d^2\mu^6/36\pi^4$, $c_S^2=1/5$ and $\delta=2/15$.

The considered examples demonstrate that local vector and diquark interaction channels break the asymptotically conformal behavior of quark matter.
Thus, nonlocal treatment of the quark-quark interactions in the NJL model is crucial for reaching the conformal limit of QCD.

\section{Neutron stars with CFLL quark cores}
\label{sec4}

Quark deconfinement significantly impacts the phenomenology of NSs.
The most common way of considering this scenario corresponds to constructing a two-phase EoS of the NS matter by matching within the Maxwell construction independent quark and hadron EoSs.
This scheme does not address the microscopic mechanism of conversion of hadrons to quarks.
While this mechanism lays beyond the scope of the present paper, we consider the recently proposed picture when BEC condensation of sexaquarks causes a mechanical instability of the NS matter against the gravitational compression and leads to a responseless density pile up followed by the Mott dissociation of hadrons and deconfinement of quarks \cite{Blaschke:2022knl,Shahrbaf:2022upc}. 
It is worth mentioning that the nonlocal NJL model developed in Section \ref{sec3} is not limited to this picture, which is considered for the sake of illustration of the possible mechanism of quark deconfinement in NSs.

Under the color gauge group diquarks transform as antitriples, thus, making the QCD interaction between them analogous to the one between antiquarks.
This potentially allows formation and BEC of a deeply bound sexaquark state of three CFL diquarks qualitatively different from the dihyperon state of the same flavor content \cite{Jaffe:1976yi}.
Unlike the dihyperon state, the sexaquark is not yet studied in lQCD.
Therefore we rely on the results obtained with QCD sum rules \cite{Azizi:2019xla} and a constituent quark model \cite{Buccella:2020mxi}, which predict $m_S=1180^{+40}_{-28}$ MeV and $m_S=1880\pm3$ MeV, respectively.
Despite being significantly different, these results support stability of sexaquarks with respect to its strong and weak decays.
Following Refs. \cite{Blaschke:2022knl,Shahrbaf:2022upc}, we prefer a conservative estimate of the sexaquark mass as the upper limit of the threshold with respect to the dominant weak process being the decay to electron, proton and $\Lambda$-baryion, i.e. $m_S=m_e+m_p+m_\Lambda=2054$ MeV.
Such sexaquark can decay only in a doubly-weak channel, which makes its lifetime compatible to the age of the Universe \cite{Farrar:2003qy}, so providing the effective stability on the timescales corresponding to NSs.
For the sake of simplicity we also ignore medium modification of the sexaquark mass caused by the Pauli blocking among the constituent quarks.
Accounting for such a dependence, however, would not drastically change the considered scenario.
It only leads to a moderate density dependence of $m_S$ and insignificantly increases ability of the sexaquark enriched NS matter to resist the gravitational compression \cite{Shahrbaf:2022upc}.

The described sexaquark is included to the hadron part of the hybrid quark-hadron EoS used below within the DD2npY-T model supplemented with a crust EoS \cite{Shahrbaf:2022upc}.
It also includes nucleons and hyperons, is consistent with the low density constraint of the chiral effective field theory (CEFT) \cite{Kruger:2013kua} and reproduces the main properties of the nuclear matter ground state.
At $m_S=2054$ MeV the BEC of sexaquarks onsets at the baryon density $n_B=0.25~{\rm fm}^{-3}$ and the energy density $\varepsilon=245~{\rm MeV~fm}^{-3}$.
The corresponding chemiocal potential $\mu_B=m_S/2$ is defined by the condition of the BEC of sexaquarks \cite{Blaschke:2022knl,Shahrbaf:2022upc}.

Further increase of density of the hadron matter occurs due to populating the BEC of sexaquarks, which leaves the pressure and chemical potential unchanged \cite{Shahrbaf:2022upc}.
Thus, the BEC of sexaquarks causes mechanical instability of hadron matter being compressed in the gravitational field of NSs unless the sexaquarks and other hadrons dissociate to quarks.
The flavor content of sexaquarks $uuddss$ coincides with the one of three diquarks $ud$, $ds$ and $su$.
Dissociation of sexaquarks causes formation of a color-superconducting three-flavor state of quark matter dominated by the above diquarks, i.e. the CFLL quark matter \cite{Blaschke:2022knl}.
This state exists in thermodynamic equilibrium with the nuclear matter, i.e. their pressures and chemical potentials coincide. 
Thus, within the considered picture quark deconfinement can be modeled by matching quark and hadron EoSs by means of the Maxwell construction at $\mu_B=m_S/2$.
This simultaneously respects the mentioned conditions of the BEC of sexaquarks and thermodynamic equilibrium.
 
Providing the Maxwell matching of the EoSs of CFLL quark matter with the DD2npY-T one of hadron matter at $\mu_B=m_S/2$ requires adjusting the vector and diquark couplings of the nonlocal NJL model.
Conventionally, below we treat $\eta_V$ as a single independent parameter of the model and adjust $\eta_D$ according to the above requirement.
It is known that increasing the vector coupling increases the onset density of quark matter, while the diquark one causes the opposite effect (see the corresponding discussion e.g. in Refs. \cite{Baym:2017whm,Ivanytskyi:2022oxv,Ivanytskyi:2022bjc}).
Therefore, having a fixed value of the chemical potential of quark deconfinement defined by the sexaquark mass requires $\eta_D$ to be a growing function of $\eta_V$.
This behavior is seen from Table \ref{table1}, which summarizes the hybrid EoSs used below.
Since the considered scenario assumes deconfinement of a color-superconducting CFLL quark matter, the diquark coupling obtained from the requirement that quark-hadron transition onsets at the BEC of sexaquarks should provide a finite diquark pairing gap found as a solution of Eq. (\ref{XXIII}) at $\mu_B=m_S/2$.
Within the considered scenario transition from hadron to quark matter in the general case is of the first order.
It is indicated by a finite jump of the baryon density.
Requiring this density jump across the transition from nuclear to quark matter to be non-negative, we limit the ranges of the vector and diquark couplings from above.
The vector and diquark couplings producing the EoSs summarized in Table \ref{table1} respect these two requirements.
They also assure the absence of the vacuum color superconductivity since they are always below the critical value $\eta_D=0.765$.

\begin{figure}[t]
\label{fig1}
\includegraphics[width=1\columnwidth]{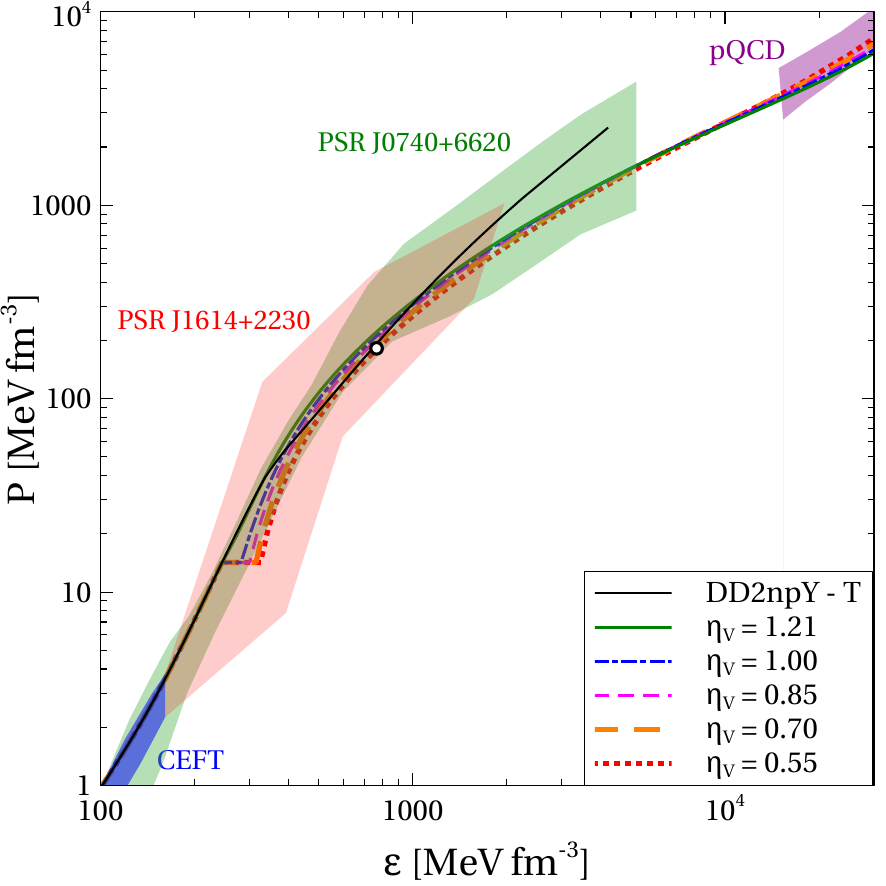}
\caption{Hybrid EoS of cold electrically neutral quark-hadron matter at $\beta$-equilibrium in the plane of pressure $p$ versus energy density $\varepsilon$. 
The curve corresponding to hadron matter is obtained without sexaquarks.
The empty white circle indicates onset of hyperons in pure hadron matter.
The calculations are performed for the sexaquark mass $m_S=2054$ MeV.
The nuclear matter \cite{Kruger:2013kua}, astrophysical \cite{Miller:2021qha,Hebeler:2013nza} and perturbative QCD \cite{Fraga:2013qra} constraints represented by the shaded areas are discussed in the text.}
\end{figure}

In order to model NSs we construct a family of hybrid quark-hadron EoSs based on the procedure described above.
The resulting EoSs in the plane of pressure versus energy density are shown in Fig. \ref{fig1}.
They agree well with the constraints extracted from the observational data of the pulsars PSR J1614+2230 \cite{Hebeler:2013nza} and PSR J0740+6620 \cite{Miller:2021qha}.
An insignificant tension with the PSR J0740+6620 constraint is due to the absence of a first order phase transition in the analysis of Ref. \cite{Miller:2021qha}, while the corresponding observational data are fully reproduced within the present model.
The flat parts of the depicted curves correspond to the mixed quark-hadron phase where the BEC of sexaquarks exists.
As is seen, increasing the vector coupling stiffens the quark branch of the EoS and simultaneously reduces the density jump across the quark-hadron transition.
The maximal considered value of $\eta_V=1.21$ corresponds to vanishing of this jump.
At sufficiently large values of this parameter the quark branch of the hybrid EoS is stiffer than the DD2npY-T EoS at the energy densities above $400~{\rm MeV~fm}^{-3}$.
As is shown below, this allows the hybrid EoSs to reach the maximum NS masses above the value $2.1~{\rm M}_\odot$ corresponding to the DD2npY-T EoS.
It is also seen from Fig. \ref{fig1} that the developed EoSs are consistent with the constraint of perturbative QCD \cite{Kurkela:2009gj,Fraga:2013qra,Gorda:2018gpy,Fernandez:2021jfr}.
This is provided by the nonlocal treatment of quark interactions allowing the present model to asymptotically reach the conformal limit of QCD.

\begin{figure}[t!]
\label{fig2}
\includegraphics[width=1\columnwidth]{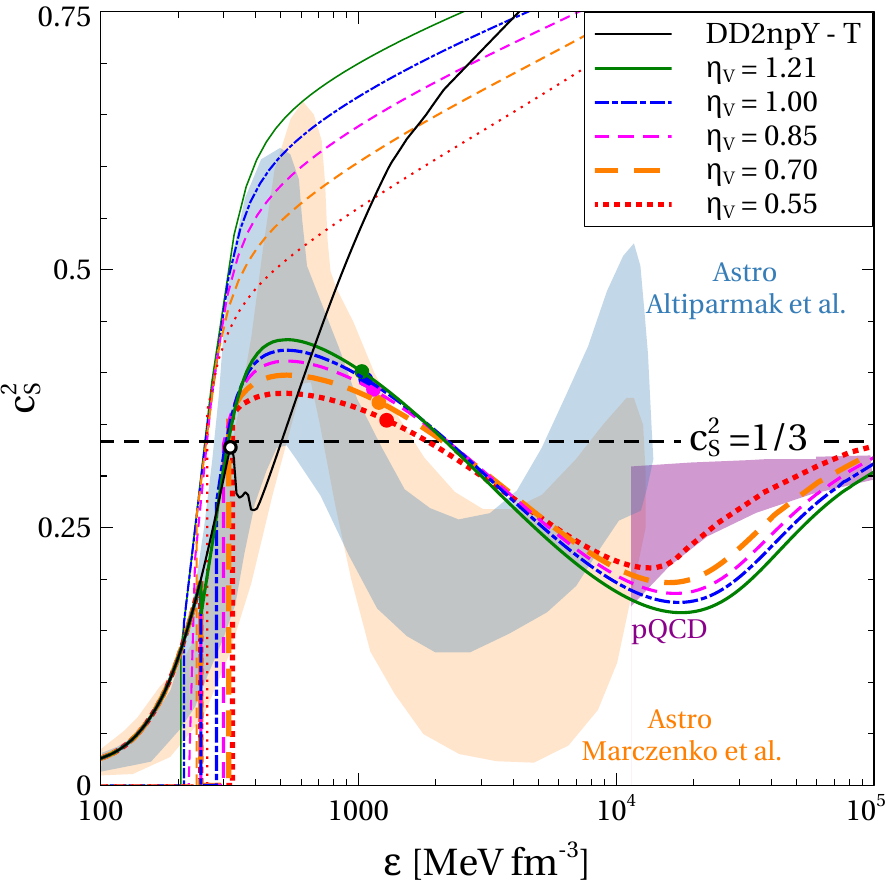}\\
\includegraphics[width=1\columnwidth]{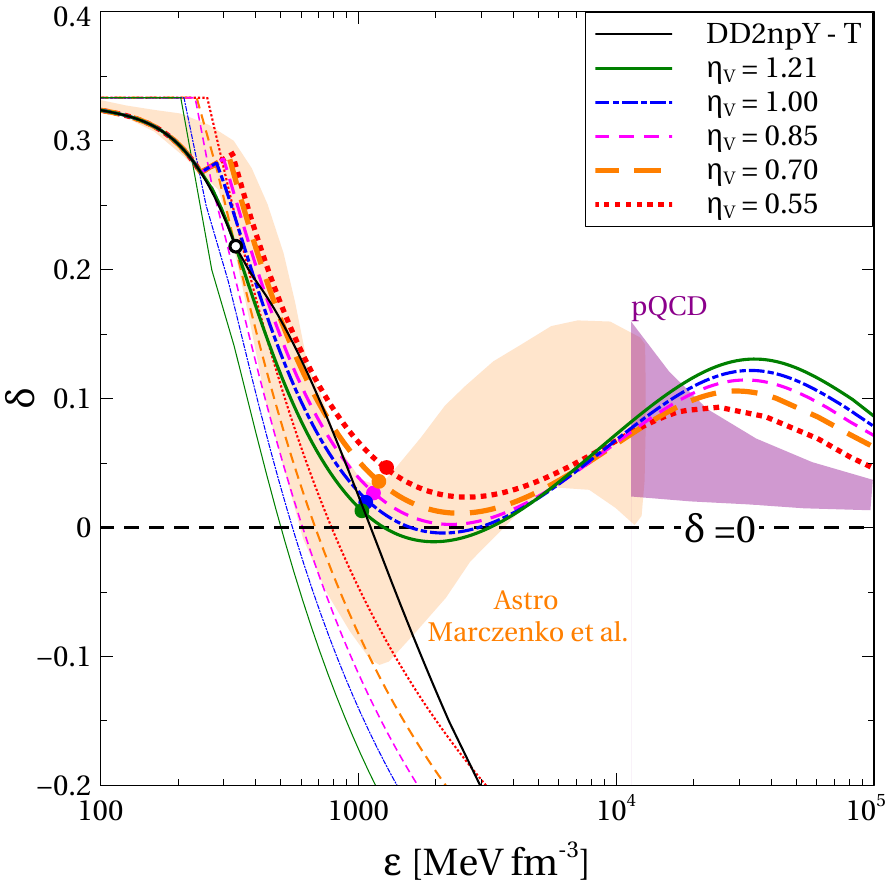}
\caption{Speed of sound $c_S^2$ (upper panel) and dimensionless interaction measure $\delta$ (lower panel) as functions of energy density $\varepsilon$ obtained for the EoSs shown in Fig. \ref{fig1} (thick curves). 
The thin curves are obtained for pure quark matter within the NJL model with local vector interaction from Ref. \cite{Blaschke:2022egm} at the same couplings as in the case of the thick curves. 
The empty white circles indicate onset of hyperons in pure hadron matter.
The filled colored circles represent the maximum baryon densities in the centers of the heaviest NSs with the corresponding EoSs.
The black dashed lines represent the conformal values $c_S^2=1/3$ and $\delta=0$.
The astrophysical \cite{Altiparmak:2022bke,Marczenko:2022jhl} and perturbative QCD \cite{Fraga:2013qra} constraints represented by the shaded areas are discussed in the text.}
\end{figure}

The speed of sound and dimensionless interaction measure of the present model are shown in Fig. \ref{fig2}.
The hadron branch of the hybrid EoS demonstrates monotonously growing $c_S^2$ and decreasing $\delta$.
In the purely hadronic case $c_S^2$ is nonmonotonous due to the onset of hyperons.
In the mixed phase region the speed of sound vanishes due to the constant pressure while the dimensionless interaction measure exhibits a non-smooth behavior caused by a discontinuous density jump.
Note that at the limiting value of the vector coupling $\eta_V=1.21$ the first order quark-hadron transition degenerates to the second order one indicated by the vanishing density jump.
In this case the region with vanishing $c_S^2$ disappears, while preserving a discontinuity of this quantity. 
At this value of the vector coupling $\delta$ is a smooth function of $\varepsilon$.
The quark branch of the hybrid EoS exhibits a nonmonotonous behavior of the speed of sound and dimensionless interaction measure.
Right after the quark deconfinement $c_S^2$ increases and $\delta$ decreases with growing energy density. 
This signals stiffening of the EoS caused by the quark deconfinement.
The larger the vector coupling, the stiffer the quark branch of the EoS.
This is manifested by increasing $c_S^2$ and decreasing $\delta$ at growing $\eta_V$.
The speed of sound develops a peak at $\varepsilon\simeq500~{\rm MeV/fm}^3$. 
Thus, $c_S^2$ reaches its maximal value in the quark matter phase,

As is seen from Fig. \ref{fig2}, the maximum of $c_S^2$ does not coincide with the minimum of $\delta$ at $\varepsilon\simeq2100~{\rm MeV/fm}^3$.
It is remarkable that this minimum corresponds to very small $\delta$ not exceeding $0.05$ by the absolute value.
In addition, the speed of sound is very close to $1/3$ in that density range.
In other words, the speed of sound and dimensioneless interaction measure of the CFLL quark matter simultaneously reach nearly conformal values in a non-perturbative range of the energy density around $\varepsilon=2100~{\rm MeV/fm}^3$.
This value of $\varepsilon$ significantly exceeds the maximal energy densities $\varepsilon_{\rm max}$ reached in the centers of the heaviest NSs with the CFLL quark matter cores. For the considered values of the vector coupling this maximal energy density is given in Table \ref{table1}.

Precursors of such a behavior of $c_S^2$ and $\delta$ in the centers of the heaviest NSs were discussed within the context of approximately restored conformal symmetry \cite{Fujimoto:2022ohj,Annala:2023cwx}.
At the same time, the conformality of a theory assumes that the corresponding action is scale invariant.
This implies an absence of dimensionful parameters in the action and leads to $p\propto\varepsilon\propto\mu^4$ due to the dimensionality reason.
In this case $c_S^2=1/3$ and $\delta=0$ follow as a consequence of scale invariance.
However, the inverse statement that $c_S^2=1/3$ and $\delta=0$ at a certain density imply the scale invariance is not correct in the general case.
A simple counterexample is given by the model of ultraheavy quarks with local vector repulsion and a constant negative bag-like pressure shift.
This model obviously breaks the scale invariance by three dimensionful parameters, i.e. by the quark mass, the vector coupling and the bag constant.
The interested readers can find the corresponding consideration in Appendix \ref{secApp}.
At the same time, at the non-perturbative regimes the CFLL quark matter is characterized by the effective quark mass, vector field amplitude and pairing gap.
These dynamically generated parameters are not negligible compared to the quark chemical potential and break the scale invariance of the effective action.
For example, at $\eta_V=1$ and $\varepsilon=2100~\rm MeV/fm^3$ one gets $\omega/\mu=0.58$, while the speed of sound and dimensionless interaction measure are close to the conformal values.
Therefore, the question about an approximately conformal behavior of quark matter at non-perturbative densities can not be addressed based just on the analysis of its EoS and requires an independent consideration of the effective action and relevant microscopic quantities.  
Such a study is beyond the scope of the present paper and should be performed separately.

Fig. \ref{fig2} also demonstrates that the scenario of the CFLL quark matter in NS agrees with the constraints on the speed of sound and dimensionless interaction measure at the energy densities reached in NSs.
These constraints were extracted from the astrophysical data based on the model agnostic statistical analysis at the 94 \% confidence level \cite{Altiparmak:2022bke,Marczenko:2022jhl}.
The mentioned agreement is rather remarkable since unlike the present work the analysis from Refs. \cite{Altiparmak:2022bke,Marczenko:2022jhl} does not explicitly include the possibility of a first order phase transition.
A possible phenomenological explanation may be related to an early quark deconfinement at an onset mass below the masses of the analyzed NSs.
In this case all these astrophysical objects lay on the quark branch of the mass-radius relation and, thus, are weakly sensitive to the details of the deconfinement phase transition.
It is also seen from Fig. \ref{fig2} that the speed of sound of the nonlocal NJL model is consistent with the perturbative QCD predictions \cite{Kurkela:2009gj,Fraga:2013qra,Gorda:2018gpy,Fernandez:2021jfr} at the energy densities above $\rm 10^5~MeV/fm^3$. 
In the case of the dimensionless interaction measure the agreement requires even larger $\varepsilon$.
As discussed in Section \ref{sec3}, this is due to the impossibility to adjust the formfactor according to the perturbative QCD results on the behavior of $c_S^2$ and $\delta$ at asymptotically high densities.  
A better agreement can be reached by a piecewise $g_{\bf k}$ being Gaussian at small momenta and behaving as $g_{\bf k}\propto\alpha_s/{\bf k}^2$ at large momenta \cite{Hell:2011ic,Kashiwa:2011td}.

As was shown in Section \ref{sec2}, the effective quark mass, vector field amplitude and pairing gap are negligible at high densities.
This insures the conformal behavior of the CFLL quark matter with the speed of sound and dimensionless interaction measure assymptoticlly converging to the conformal values $c_S^2=1/3$ and $\delta=0$.
Fig. \ref{fig2} demonstrates this.
It is also seen that convergence to the conformal limit is delayed by increasing the vector coupling.
For the Gaussian formfactor and a finite current quark mass $c_S^2\simeq1/3-m^2/3\mu^2$ and $\delta\simeq m^2/3\mu^2$ at high densities, while the terms corresponding to the vector repulsion are subdominant corrections.
It follows from Eq. (\ref{XXXII}) that they can be neglected if $\mu^3 g_\mu\ll-m^2\Lambda^2/\omega_\infty$. 
This condition is strongly dependent on the vector coupling appearing through the asymptotic value of the vector field. 
For example, at $\eta_V=0.55$ and $\eta_V=1.00$ this requires $\mu_B\gg6439$ MeV, $\varepsilon\gg5.10\cdot10^7~\rm MeV/fm^3$ and $\mu_B\gg6586$ MeV, $\varepsilon\gg5.58\cdot10^7~{\rm MeV/fm}^3$, respectively.
At smaller chemical potentials and energy densities vector repulsion significantly affects the behavior of speed of sound and dimensionless interaction measure.

Comparing the speed of sound and dimensionless interaction measure of the present model and the NJL model with local vector repulsion from Ref. \cite{Blaschke:2022knl} allows us to analyze the phenomenological influence of the nonlocal treatment of the mentioned interaction channel.
Fig. \ref{fig2} shows $c_S^2$ and $\delta$ of pure quark matter obtained for the model with the local vector interaction. 
The first order phase transition from chirally broken normal quark matter to partially chirally symmetric color-superconducting quark matter is indicated in in Fig. \ref{fig2} by discontinuous jumps of $c_S^2$ and kinks of $\delta$.
In the case of local treatment of the vector repulsion the NJL model produces rather stiff EoS of quark matter.
This leads to large values of the speed of sound always exceeding the one corresponding to the nonlocal NJL model.
As is seen, unlike the nonlocal NJL model, $c_S^2$ monotonously grows asymptotically reaching the speed of light (see the discussion in Section \ref{sec3}).
As argued above, the local treatment of the vector repulsion does not allow the NJL model to reach the conformal limit.
The same conclusion can be drawn from the behavior of the dimensionless interaction measure.
At the local vector repulsion it monotonously decreases saturating to the nonconformal value $-2/3$.
It is also worth mentioning that in the case of the local treatment of the vector repulsion the quark matter EoS does not exhibit a region where $c_S^2$ and $\delta$ simultaneously attain nearly conformal values.

\begin{figure}[t]
\label{fig3}
\includegraphics[width=1\columnwidth]{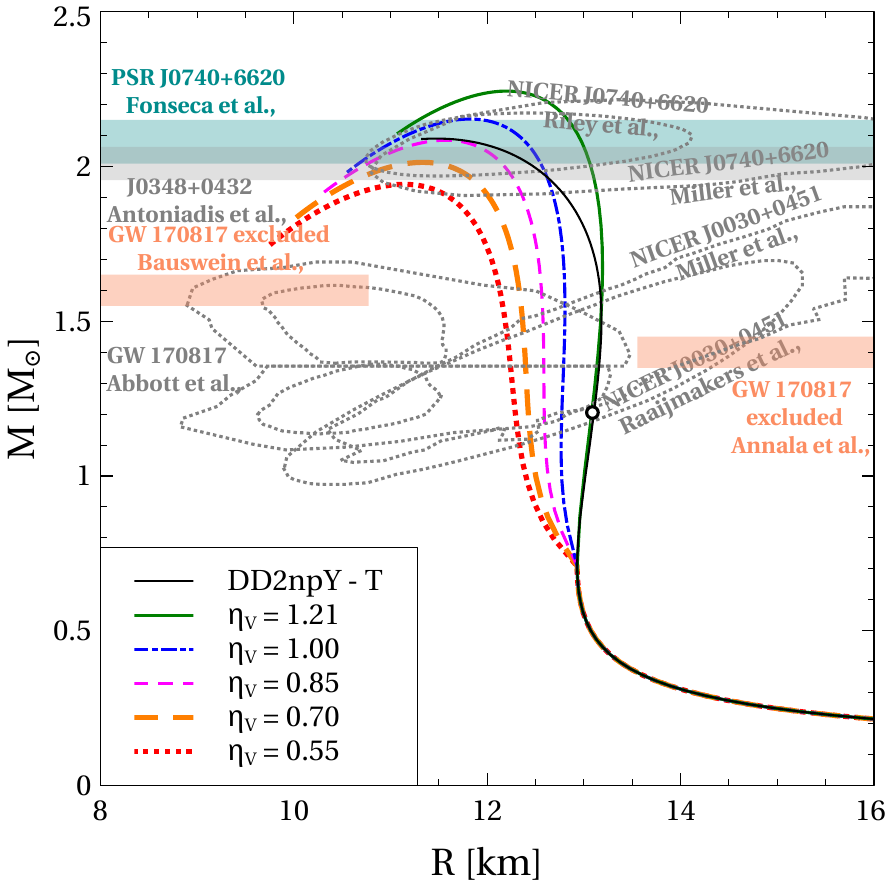}
\caption{Mass-radius relation obtained for the EoSs shown in Fig. \ref{fig1}. 
The empty white circle indicates onset of hyperons in pure hadron matter.
The represented observational constraints from Refs. \cite{Antoniadis:2013pzd,Fonseca:2021wxt,Miller:2021qha,Riley:2021pdl,Miller:2019cac,Riley:2019yda,Raaijmakers:2019qny,LIGOScientific:2018cki,Bauswein:2017vtn,Annala:2017llu} are discussed in the text.}
\end{figure}

The constructed family of EoSs allows us to solve the problem of relativistic hydrostatics constituted by the Tolman-Oppenheimer-Volkoff (TOV) equations with the proper boundary conditions \cite{Tolman:1939jz,Oppenheimer:1939ne} and construct the relation between the mass $\rm M$ and radius $\rm R$ of NSs with CFLL cores.
Fig. \ref{fig3} depicts this relation.
The chosen value of the sexaquark mass leads to the onset mass of quark deconfinement $0.72~{\rm M}_\odot$, which is before the hyperon onset mass $1.21~{\rm M}_\odot$ in the case of purely hadronic NSs.
Increasing the vector coupling stiffens the quark branch of the EoS and yields larger maximum masses of NSs and their radii.
At the same time, the early quark deconfinement and non-vanishing density jump at the quark-hadron transition ($\eta_V<1.21$) make the radius of a canonical NSs with the CFLL quark core smaller than in the purely hadronic case, while keeping the maximum NS mass large.
This improves agreement with the present observational constraints.
As is seen from Fig. \ref{fig3} and Table \ref{table1}, the considered scenario of quark deconfinement via the BEC of sexaquarks is consistent with the lower limit of the TOV maximum mass measured in a binary white dwarf-pulsar PSR J0348+0432 system \cite{Antoniadis:2013pzd} and extracted from the continued timing observations of the pulsar PSR J0740+6620 \cite{Fonseca:2021wxt}.
It also agrees with the results of the Bayesian analysis of
the observational data from the pulsars PSR J0740+6620
\cite{Miller:2021qha,Riley:2021pdl} and PSR J0030+0451 \cite{Miller:2019cac,Riley:2019yda,Raaijmakers:2019qny}.
The constraints obtained from the gravitational wave signal from the NS merger GW170817 \cite{LIGOScientific:2018cki,Bauswein:2017vtn,Annala:2017llu} are also respected.

\begin{figure}[t]
\label{fig4}
\includegraphics[width=1\columnwidth]{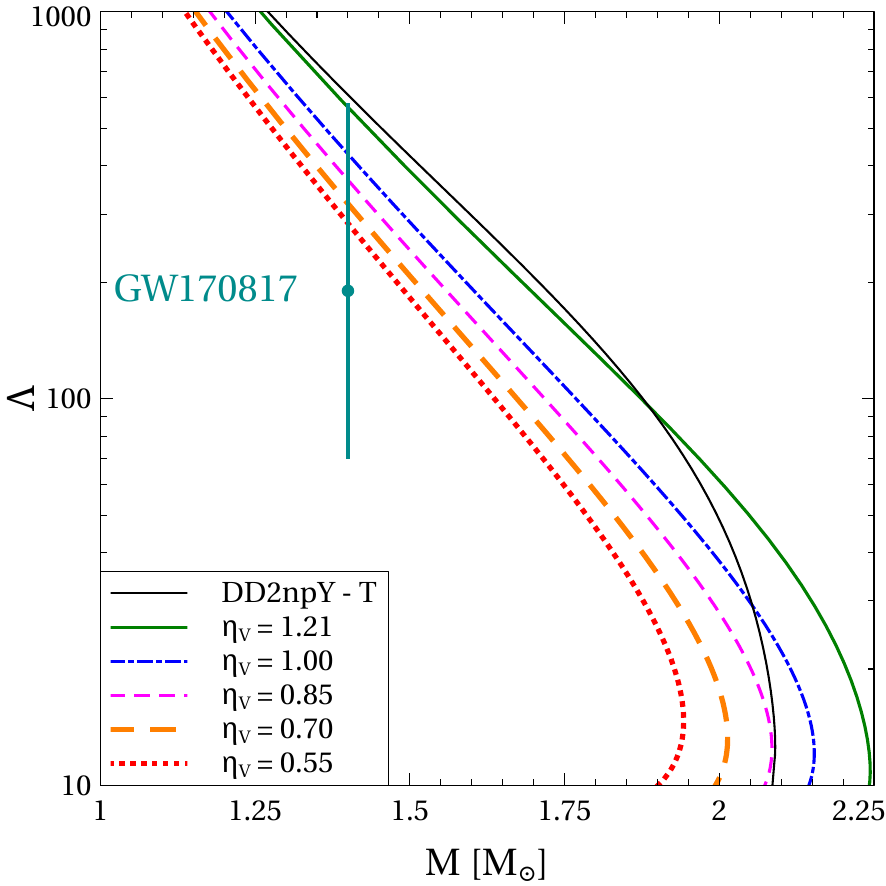}
\caption{Tidal deformability $\Lambda$ as a function of the NS mass $\rm M$ obtained for the EoSs shown in Fig. \ref{fig1}. 
The observational constraint extracted from the gravitational wave signal GW170817 \cite{LIGOScientific:2018cki} is discussed in the text.}
\end{figure}

The tidal deformability of NSs $\Lambda$ is another observable, which is sensitive to the NS matter EoS.
Dependence of this quantity on the NS mass is shown in Fig. \ref{fig4}.
Increasing the vector coupling leads to larger tidal deformability since $\Lambda\propto\rm R^5$ and, as it was discussed before, $\rm R$ grows with $\eta_V$.
We confront the developed model to the observational constraint on the tidal deformability of a $1.4~\rm M_\odot$ NS $\Lambda_{1.4}$ extracted from the gravitational wave signal GW170818. 
The multimessenger data analysis suggests $70<\Lambda_{1.4}<590$ \cite{LIGOScientific:2018cki}.
At the maximal considered $\eta_V$ corresponding to the vanishing density jump at the quark-hadron transition this constraint is fulfilled marginally.
At smaller values of the vector coupling the present model provides a good agreement with the mentioned constraint.
At the same the used hadronic EoS contradicts this constraint.
Thus, we conclude that the scenario of the CFLL matter deconfinement via the BEC of sexaquarks is consistent with the limitation on the tidal deformability of a canonical NS (see Table \ref{table1}).

\begin{table}[t]
\centering
\begin{tabular}{|c|c|c|c|c|c||c|c|c|c|c|c|c}
\hline    
EoS & $\Delta^*$ [MeV] & $\rm M_{\rm max}~[M_\odot]$ & $\rm R_{1.4}$ [km] & $\Lambda_{1.4}$ \\ \hline
set I b   & 100 & 2.20 & 13.04 & 623 \\ \hline
set II b  & 128 & 2.05 & 12.32 & 402 \\ \hline
set III b & 202 & 1.69 &  9.96 & 81  \\ \hline
\end{tabular}
\caption{Pairing gap parameters $\Delta^*$, maximum masses $\rm M_{\rm max}$ as well as radii $\rm R_{1.4}$ and tidal deformabilities $\Lambda_{1.4}$ of a canonical NS corresponding to the hybrid EoSs from Ref. \cite{Blaschke:2022egm}.
According to Ref. \cite{Blaschke:2022egm}, the EoSs are referred to as set I b, set II b and set III b.
The quark onset mass is $0.84~\rm M_\odot$ for all the EoSs.
The data are extracted from Table I of Ref. \cite{Blaschke:2022egm}.} 
\label{table2}
\end{table}

As it was shown in Section \ref{sec3}, nonlocality of the quark-quark interaction is crucial for reaching the conformal limit of QCD.
It is also important for the phenomenology of NSs with quark cores.
To demonstrate this we compare the results of this work to the ones obtained within the NJL model with local vector interaction from Ref. \cite{Blaschke:2022egm}. 
More specifically, we consider the mass-radius and tidal deformability-mass relations from Ref. \cite{Blaschke:2022egm}, which correspond to the quark onset mass $0.84~\rm M_\odot$ being close to the value $0.72~\rm M_\odot$ used in this paper.
These relations were obtained for hybrid quark-hadron EoSs, which can be classified by the value of a constant pairing gap parameter $\Delta^*$ controlled by the couplings of the NJL model.
Note, the Maxwell crossing of the stiff quark EoS obtained within the NJL model with the local vector repulsion and the hadron EoS used in Ref. \cite{Blaschke:2022egm} was provided by introducing a necessary bag-like pressure shift.
The parameter $\Delta^*$ was derived within the NJL model considered in Ref. \cite{Blaschke:2022egm} by mapping the corresponding EoS to the simple phenomenological parameterization by Alford, Braby, Paris and Reddy \cite{Alford:2004pf}.
Table \ref{table2} summarises some parameters of NSs obtained using those EoSs.
As is seen, the NJL model with local vector interaction from Ref. \cite{Blaschke:2022egm} can simultaneously describe the mentioned observational constraints on $\rm M_{\rm max}$ and $\Lambda_{1.4}$ only for a narrow range of the pairing gap parameter.
Too small $\Delta^*$ lead to too large tidal deformabilities, while too large values of this parameter do not support sufficiently large maximum masses of NSs.
The tension with the tidal deformability is due to an redundant stiffness of the quark EoS with local vector repulsion and, consequently, large radii of a canonical NS.
This redundant stiffness can be read of from Fig. \ref{fig2}.
Compensating it requires strengthening the diquark pairing manifested by larger values of the pairing gap parameter. 
However, increasing $\Delta^*$ so that $\Lambda_{1.4}$ is reproduced within the NJL model with local vector repulsion sizably softens the EoS creating a tension with large maximum masses.
This problem does not appear if the vector repulsion between quarks is treated nonlocally so that the EoS of quark matter is not redundantly stiff.
Thus, nonlocality of quark-quark interactions is important for the NS phenomenology since it supports the existence of heavy NSs with not too large tidal deformabilities.

\section{Conclusions}
\label{concl}

In the work we for the first time have derived a version of the three-flavor nonlocal NJL model with the scalar attractive, vector repulsive and diquark pairing interaction channels.
The nonlocality of the quark-quark interactions was introduced within the separable approximation.
It is controlled by the momentum dependent formfactor chosen in the Gaussian form.
The model was considered in the case when three quark flavors are degenerate in mass, which corresponds to the CFLL quark matter.
This simplification allows obtaining the single particle energies analytically. 
The model was applied to construct the EoS of cold and dense quark matter. 
Analysis of its high density asymptotics suggests that in the considered regime color superconductivity is a ground state of quark matter in agreement with the perturbative QCD conclusion.
Reaching the conformal limit at high densities as $c_S^2\rightarrow1/3-0$ and $\delta\rightarrow1/3-c_S^2$ also respects the perturbative QCD predictions.
These features of the model are shown to be provided by the nonlocal character of the quark-quark interactions.
At the same time, none of the formfactors guaranteeing the renormalizability of the model allows it to reproduce the asymptotic behavior of speed of sound predicted by the two-loop perturbative results.
This manifests an inherent non-perturbative character of the NJL model.

The developed model was applied to the scenario when the BEC of a color-spin-flavor singlet sexaquark state triggers deconfinement of the CFLL quark matter in NSs.
This transition occurs due to a mechanical instability of the sexaquark enriched cold hadron matter.
In the presence of the gravitational field of NS such matter responselessly compresses until the Mott dissociation of nucleons and sexaquarks, which deconfines the CFLL quark matter.
The key parameter of this scenario defining the onsets of the BEC of sexaquarks and, consequently, the CFLL quark matter is the sexaquark mass. 
Motivated by the results obtained with QCD sum rules and a constituent quark model and conservatively estimated as the upper limit of the dominant weak decay threshold, the used value of the sexaquark mass corresponds to an early quark deconfinement in NSs with the onset mass $0.72~{\rm M}_\odot$.
The proposed scenario is shown to be consistent with all the available observational constraints on the mass-radius relation and tidal deformability of NSs.
It simultaneously provides rather small radius of a canonical NS and the maximum NS mass exceeding the one obtained within the purely hadronic scenario.
The high maximum masses of hybrid NSs are due to stiffening of the quark matter EoS signaled by the speed of sound peak right after the quark-hadron transition. 

Interestingly, the developed EoS of the NS matter also demonstrates a non-perturbative range of energy densities around $2100~\rm GeV/fm^3$, where the speed of sound and dimensionless interaction measure simultaneously get very close to the conformal values.
However, this range exceeds the maximal energy densities reached in the centers of the heaviest NSs by 50 \% - 100 \%.
Thus, it is not clear whether the NS matter can be nearly conformal.
It is important to stress, the question about a nearly conformal behavior of quark matter at non-perturbative regimes requires a separate study of the scale invariance of the effective action and relevant microscopic quantities.

An aspect missing in the present work is the strange quark mass.
Treating it independently from the masses of light quarks will enable a more accurate determination of the model parameters including the vector coupling by fitting them to the vacuum masses of the corresponding mesons.
We also argue that the effects of the physical current mass of strange quarks do not break the CFL pairing pattern above the quark-hadron transition and can be accounted for perturbatively, which makes the CFLL picture a good approximation to cold and dense QCD matter.
Another potential improvement of the model is related to using piecewise formfactors to reach a better agreement with the perturbative QCD.
Finally, extending the analysis to finite temperatures and isospin asymmetries are important for modelling the QCD phase diagram and such astrophysical applications as NS mergers and supernovae explosions.

\vspace{0.5cm}
\section*{acknowledgments}
O.I. acknowledges fruitful discussions with David Blaschke and his assistance as well as valuable help of Pasi Huovinen.
The analysis of the dimensionless interaction measure was inspired by comments of Kenji Fukushima.
This work was performed within the program Excellence Initiative--Research University of the University of Wrocław of the Ministry of Education and Science and received funding from the Polish National Science Center under grant No. 2021/43/P/ST2/03319.  
This work was supported by INCD and FCT I.P. under the project Advanced Computing Project 2023.10526.CPCA.A2, platform Cirrus.

\begin{appendix}
\section{Ultraheavy quarks with local vector repulsion}
\label{secApp}

At finite densities the Fermi momentum of ultraheavy quarks $k_F=(\pi^2 n_B)^{1/3}$ is negligible compared to their mass $m$. 
This justifies ignoring the corresponding contribution to the energy density, which can be written as
\begin{eqnarray}
    \label{AI}
    \varepsilon=3mn_B+an_B^2+B.
\end{eqnarray}
The first term in this expression stands for the single particle contribution of ultraheavy quarks with the approximate dispersion relation $\epsilon_{\bf k}\simeq m$, the second one accounts for the effects of local vector repulsion controlled by the coupling $a$, while $B$ is a constant energy density shift.
Using the thermodynamic identities $\varepsilon=\mu_B n_B-p$ and $\mu_B=\partial\varepsilon/\partial n_B$, the pressure of the considered model arrives at
\begin{eqnarray}
    \label{AII}
    p=an_B^2-B.
\end{eqnarray}
Eqs. (\ref{AI}) and (\ref{AII}) allow us to express
\begin{eqnarray}
    \label{AIII}
    \delta&=&\frac{3mn_B-2an_B^2+4B}{3mn_B+an_B^2+B},\\
    \label{AIV}
    c_S^2&=&\frac{2an_B}{3m+2an_B}.
\end{eqnarray}
At $B=-9m^2/32a$ these  expressions simultaneously yield $\delta=0$ and $c_S^2=1/3$ at $n_B=3m/4a$. 
The consistency of the consideration requires $m\gg k_F$ at this density, i.e. $m^2a\gg3\pi^2/4$.
This condition is respected by ultraheavy quarks with an arbitrary nonvanishing vector repulsion.
Thus, the considered model breaks the scale invariance by large $m$ but yields the conformal values of the speed of sound and dimensionless interaction measure at a certain density.
Note, $a$ and $B$ can be either small or large.
Thus, in the general case they also break the scale invariance of the considered model.

\end{appendix}

\bibliography{bibliography}

\end{document}